\documentclass[prd,nofootinbib,showpacs,showkeys,twocolumn,a4paper]{revtex4}
\pdfoutput=1

\usepackage{graphicx}
\usepackage{amsmath,amssymb}
\usepackage{amsfonts}
\usepackage{latexsym}
\usepackage{mathrsfs}
\usepackage{color}
\usepackage{slashed}
\usepackage{float}
\usepackage{feyn}
\usepackage[all]{xy}
\usepackage{feyn}
\usepackage{hyperref}
\usepackage{dsfont}
\usepackage{placeins}
\usepackage[justification=raggedright ,singlelinecheck=false,font=small]{caption,subfig}

\allowdisplaybreaks

\definecolor{blue}{rgb}{0,0,0.5}
\definecolor{lightblue}{rgb}{0,0,1}
\definecolor{red}{rgb}{0.5,0,0}
\definecolor{lightred}{rgb}{1,0.5,0}
\definecolor{green}{rgb}{0,0.5,0}
\definecolor{darkgreen}{rgb}{0.0,0.3,0.0}
\definecolor{orange}{rgb}{1,0.4,0}
\definecolor{grey}{rgb}{0.5,0.5,0.5}
 
\providecommand{\eqn}{Eq.~}

\providecommand{\fig}{Fig.~}

\providecommand{\sect}{Sec.~}
\providecommand{\app}{Appendix~}

\newcommand{\printifnonempty}[2]{\if\relax\detokenize{#1}\relax\else#2\fi}
\newcommand{\alter}[5]{%
  \long\def\temp{#3}%
  \long\def\accept{#5}%
  \ifx\temp\accept
    {#1}
  \else
    {\textcolor{#4}{\printifnonempty{#1}{{#1}}}%
    \textcolor{grey}{\printifnonempty{#2}{(#2)}}%
    \textcolor{#4}{\printifnonempty{#3}{{[#3]}}}}%
  \fi
}

\global\long\def\e{\epsilon}

\global\long\def\l{\lambda}

\newcommand{\ev}[1]{\ensuremath{\left\langle #1 %
                     \right\rangle}} 

\usepackage{feynmp}
\usepackage{bbold}
\usepackage{bbm}
\usepackage{makecell}
\usepackage{lscape}
\DeclareMathOperator{\Tr}{Tr}

\begin{document}

\pacs{miow}

\title{Neutrino Masses and Conformal Electro-Weak Symmetry Breaking}

\author{Manfred Lindner}
\email[\,]{manfred.lindner@mpi-hd.mpg.de}   

\author{Steffen Schmidt}
\email[\,]{schmidt@mpi-hd.mpg.de}     

\author{Juri Smirnov}
\email[\,]{juri.smirnov@mpi-hd.mpg.de}

\affiliation{
Max-Planck-Institut f\"ur Kernphysik, Saupfercheckweg 1, 69117 Heidelberg, Germany}

\begin{abstract}
Dimensional transmutation in classically conformal invariant theories may explain the electro-weak scale and the fact that so far nothing but the Standard Model (SM) particles have been observed. We discuss in this paper implications of this type of symmetry breaking for neutrino mass generation. 
\end{abstract}

\maketitle

\section{\label{sec:Introduction}Introduction}

A key feature of quantum field theory (QFT) is that it can not predict overall scales. 
Scale ratios are, however, calculable and this leads to the question how large ratios can be explained 
or made natural. Symmetries play here an important role. The fermions of the Standard Model (SM) 
are protected by chiral symmetry such that only logarithmic corrections occur, while the SM Higgs 
mass is unprotected which leads to the famous hierarchy problem. As a consequence one expects 
either new physics in the TeV-range or a new symmetry which also leads to new particles in the TeV-range. 
This is one of the main motivations for the LHC, but so far no new particles or interactions 
showed up. Even though there are good reasons that e.g. supersymmetric particles show up at a 
somewhat higher scale one may wonder if the fact that so far no new particle has been found points 
into some other direction. A potential role of conformal symmetry has therefore recently been discussed 
as a solution and we would like to discuss in this paper the implications for neutrino mass generation. 

It is interesting to note that the standard model of particle physics (SM) is nearly conformal invariant. 
Only the mass term of the scalar field which is responsible for the breaking of $SU(2)_L \times U(1)_Y$ 
symmetry violates conformal symmetry explicitly and all SM masses are directly proportional to this 
scale. Note that the introduction of an explicit Higgs mass term in the SM does not only break conformal 
invariance, but it also creates the hierarchy problem, namely the quadratic sensitivity of quantum corrections 
to high scales. It is therefore tempting to relate the breaking of conformal symmetry with the generation of 
the electro-weak (EW) scale by dimensional transmutation. Scale invariance is broken at the quantum level 
(i.e. it has an anomaly) even in perturbation theory \cite{Callan:1970yg}, but it has been argued that the 
protective features of conformal symmetry may not be completely destroyed \cite{Bardeen:1995kv}. Specifically 
logarithmic sensitivities still would exist, while quadratic divergencies would be absent. Various attempts 
of this type exist in the literature  \cite{Coleman:1973jx,Fatelo:1994qf,Hempfling:1996ht,Hambye:1995fr,Meissner:2006zh,Foot:2007as,Foot:2007ay,Chang:2007ki, Hambye:2007vf, Meissner:2007xv, Meissner:2009gs,Iso:2009ss,Holthausen:2009uc, Iso:2009nw,Foot:2010et, Khoze:2013uia,Kawamura:2013kua,Gretsch:2013ooa, Heikinheimo:2013fta,Gabrielli:2013hma,Carone:2013wla,Khoze:2013oga,Englert:2013gz,Farzinnia:2013pga,Abel:2013mya, Foot:2013hna,
Hill:2014mqa,Guo:2014bha,Radovcic:2014rea,Khoze:2014xha,Smirnov:2014zga,Kannike:2014mia,Chankowski:2014fva} and applications for 
the breaking of the EW symmetry have recently received more attention.

Realizing these ideas within the SM corresponds to the Coleman-Weinberg effective potential, where
$m_t < 79$~GeV would be required and where the Higgs mass would have to be $m_H \simeq 9$~GeV.
This is obviously ruled out. However, we know that the SM is incomplete, since neutrino masses must 
be included. Furthermore, there is no dark matter (DM) candidate in the SM. Phenomenologically 
successful models which employ conformal electro-weak symmetry breaking require therefore some
extension and a number of them predict also interesting DM candidates 
\cite{Farzinnia:2014xia,Guo:2014bha,Khoze:2013uia,Carone:2013wla,Ishiwata:2011aa,Cheng:2004sd,Foot:2010av,Hambye:2007vf,Radovcic:2014rea}.

In this paper we focus on neutrino masses and we argue that the dynamical generation of scales 
forbids any explicit Majorana or Dirac mass term which would otherwise be possible and expected 
for a given set of fermions. This implies that all mass terms must be dimensionless Yukawa couplings 
times one of the vacuum expectation values (VEVs) generated by the dynamical symmetry breaking. 
This clearly alters expectations for neutrino masses and we will discuss how this leads naturally to
a generic TeV scale see-saw, inverse see-saw and pseudo-Dirac scenarios.

It has been shown in \cite{Meissner:2006zh, Foot:2007ay} that extending the SM by merely right-handed 
neutrinos and an additional scalar field can result in the correct low energy phenomenology. The basic 
idea is that introducing additional scalar degrees of freedom makes the running of the couplings such that 
spontaneous symmetry breaking is possible. The additional scalar singlet gets a VEV 
and by its admixture to the Higgs a mass term is generated which can again induce EW symmetry breaking. 
This cascading symmetry breaking mechanism results in the discussed model in the correct Higgs mass 
and VEV. Thus, the EW scale appears naturally given the particle content of the model. 

The simplest model compatible with data contains a complex scalar singlet \cite{Meissner:2008gj} and 
the symmetry breaking takes place entirely in the new scalar sector, then it is transmitted via the Higgs 
portal to the SM boson. Explicit Majorana masses are not allowed and Majorana mass terms arise 
via Majorana-Yukawa couplings to the new scalar, which exemplifies nicely how neutrino mass generation
is affected. Note that this has immediate consequences for the expected Majorana mass terms. Usually,
an explicit mass is expected to have the largest possible value allowed by the symmetries of the system,
while it is now the product of the symmetry breaking hidden scalar with a TeV-scale VEV with a Yukawa 
coupling. Since the Yukawa couplings of the SM show numerically a huge range, we assume the same 
to be true more general for all Yukawa couplings and Majorana mass terms can consequently have now 
any value between zero and the symmetry breaking scale.

Motivated by this simple example we would like to discuss in this paper the changes for neutrino 
mass terms in conformally invariant theories in a more general way. We give therefore 
in \sect \ref{sec:Rules} an overview of the considered cases for the generation of neutrino masses 
within the framework of conformal theories. The consequences for the possible structure of VEVs 
are elaborated in the same paragraph. On the other hand we investigate if different conformally 
invariant neutrino mass models are possible at all with regard to the occurrence of radiative 
symmetry breaking and the correct Higgs mass. The different models are presented in \sect \ref{sec:Models} 
and are divided into two parts. The first part is based on mere extensions of the particle content of the SM, 
whereas the second part consists of theories that extend the SM gauge group by a $U(1)$ symmetry 
which separates a Hidden Sector (HS) from the SM. Different models within these parts are organized 
by their effects on the neutrino mass matrix $\mathcal{M}$. 

For neutrino masses it is in this context crucial that conformal symmetry forbids explicit mass scales 
in the classical Lagrangian. Phenomenological viable conformal EW symmetry breaking employs 
Higgs portals which connect to another sector with TeV scale dynamical mass generation. This implies
that all Dirac and Majorana fermion masses are governed by this TeV scale or by the EW scale times
some Yukawa coupling. This severely affects expectations for neutrino masses. A parameter scan for 
an effective model reveals that there are basically four phenomenological classes of theories. This 
scan is performed in \sect \ref{sec:Phenomenology}. We summarize our findings and conclude with 
a discussion in \sect \ref{sec:Conclusion}.

\section{\label{sec:Rules} Model Building Rules}

In this section we present model building rules for neutrino masses in a theory with classically conformal 
Lagrangian. Specifically we consider the following cases for neutrino masses in extensions of the standard model. 

\begin{itemize}
\item The SM can be embedded in a larger gauge group, which breaks to the required gauge group to describe the observed particle spectrum as is the case in GUT models.
\item The SM gauge group can be left unchanged and additional fields postulated.
\item A Hidden Sector (HS) with an additional symmetry group can be postulated. Resulting in the total symmetry group being a  direct product of the SM and the new sector $G(SM) \times G(HS)$.
\end{itemize}

In the following we will assume that the latter two possibilities are relevant, since the embedding of the SM in a larger gauge sector requires an additional scale of symmetry breaking which itself poses a little hierarchy problem, as in \cite{Holthausen:2009uc} where additional parameter tuning is required.  Furthermore, the additional symmetry is assumed to be global to avoid the need for anomaly cancellation at this point.

\subsection{General Conformal Building Rules}
    \label{chap:rules}

A fermion mass term is a chirality flip of the field. Therefore, we will have an incoming particle of 
one chirality, e.g. the left-handed neutrino $\nu_L$ and its antiparticle of opposite chirality as an 
outgoing particle, which is right-handed. This particle can either be its own antiparticle with a Majorana 
mass or a distinct particle with a Dirac mass. The operators in the Lagrangian have dimension three 
and thus have to be augmented by a dimension one scalar field in order to fulfil the conformal requirements. 
Thus we assume the fermions only to couple via Yukawa couplings of the form
    \begin{equation}
      \label{eq:Mass}
      \overline{\psi_L}\psi_R\varphi \; \text{ and } \; \overline{\psi_R}\psi_L\varphi \, , 
    \end{equation}
    where the $\psi$ are fermions and $\varphi$ represents a scalar.
Explicit mass terms are forbidden in the Lagrangian, i.e. any diagram like 
 \vspace*{-4mm}
    \unitlength = 1mm
    \begin{fmffile}{mass}
      \begin{eqnarray*}
        \parbox{25mm}{
        \begin{fmfgraph*}(35,20)
          \fmfleft{L}
          \fmfright{R}
          
          \fmf{plain}{L,V}
          \fmf{plain}{V,R}
          
          \fmfv{decor.shape=cross,decor.size=7}{V}
        \end{fmfgraph*}
        }
      \end{eqnarray*}
    \end{fmffile}	

 \vspace*{-4mm}
with an explicit fermion mass term (cross) is forbidden.
Yukawa couplings and mass terms which are generated via Yukawa and VEVs couplings like 
     \begin{equation}
      \label{eq:Yukawa}
      y \,\overline{\psi_L}\psi_R v_\varphi \; \text{ and } \; y\,\overline{\psi_R}\psi_L v_\varphi \, . 
    \end{equation}
are allowed:

    \vspace*{3mm}
    \begin{fmffile}{insertion}
      \begin{eqnarray*}
        \parbox{40mm}{
        \begin{fmfgraph*}(40,40)
          \fmfleft{L}
          \fmfright{R}
          \fmftop{T}
          
          \fmf{plain}{L,V}
          \fmf{plain}{V,R}
          
          \fmffreeze
          
          \fmf{dashes}{V,T}
        \end{fmfgraph*}
        }
        \hspace{3mm}
        \parbox{40mm}{
        \begin{fmfgraph*}(40,40)
          \fmfleft{L}
          \fmfright{R}
          \fmftop{T} \fmflabel{$\langle \varphi \rangle$}{T}
          
          \fmf{plain}{L,V}
          \fmf{plain}{V,R}
          
          \fmffreeze          
          
          \fmf{dashes}{T,V}
        \end{fmfgraph*}
        }
      \end{eqnarray*}
      
    \end{fmffile}
\vspace*{-6mm}
        
Each neutrino mass diagram needs an odd number of mass insertions.
Note that we work within the flavour basis, i.e. we use fields that appear in the unbroken Lagrangian.

For the scalars conformal invariance only allows couplings which connect 4 scalars, i.e. diagrams of the form
    \vspace{3mm}
    \begin{fmffile}{potential}
      \begin{eqnarray*}
        \parbox{25mm}{
        \begin{fmfgraph*}(25,20)
          \fmfleft{L1,L2}
          \fmfright{R1,R2}
          
          \fmf{dashes}{L1,V}
          \fmf{dashes}{L2,V}
          \fmf{dashes}{R1,V}
          \fmf{dashes}{R2,V}
        \end{fmfgraph*}
        }
      \end{eqnarray*}
    \end{fmffile}
    \vspace{3mm}\\

 These rules will be used throughout this work and will serve to derive rules with regard to specific neutrino mass questions.
    
     \subsection{The Weinberg Operator Case}
        We will argue that all neutrino mass diagrams, leading to a Majorana mass contribution for the active 
        neutrinos, involve at least one vacuum expectation value other than the Higgs VEV and show that this is a topological
        necessity of conformally invariant theories including upto $SU(2)$ triplet representations. \\
        To prove this we first note that any diagram has an even number of doublet scalar mass 
        insertions. This is because all diagrams generating left-handed Majorana masses have the
        left-handed doublet as the incoming and the outgoing particle, i.e. we have to start and
        end up with a doublet. If we assume that the theory has only upto $SU(2)$ triplet scalars
        and fermions, the only possibilities to connect two fermionic doublets are Yukawa couplings
        with a scalar triplet or singlet. Connecting a doublet fermion to a singlet fermion involves
        a doublet scalar. Equivalently a doublet and a triplet fermion are connected via a scalar 
        doublet. Furthermore, two fermion singlets connect to a singlet scalar, two fermion triplets
        to a singlet scalar as well and a triplet and singlet fermion to a triplet scalar
        (see Table \ref{tab:Yukawa}).
        
        \begin{table}[t]
        \begin{center}
        \scalebox{1.2}{
        \begin{tabular}{c||c|c|c}
          & $S$ & $D$ & $T$ \\
          \hline \hline 
          $S$ & $\varphi \overline{S} S$ & $\overline{D} \tilde{\phi} S$ 
          & $\Tr{[\overline{T} \Delta S]}$ \\
          \hline
          $D$ &  & $\overline{D} D^c \varphi \; , \; \overline{D} \boldsymbol{\Delta} D^c$
          & $\tilde{\phi}^\dagger \overline{T} L$ \\
          \hline
          $T$ &  &  & $\Tr{[\varphi \overline{T^c} T]}$ 
        \end{tabular}
        } 
        \caption{Possible dimension 4 Yukawa coupling terms. S, D and T denote singlet, doublet 
        and triplet fermions respectively. $\varphi$, $\phi$ and $\Delta$ denote singlet, doublet
        and triplet scalars respectively. The totally antisymmetrc coupling of three triplets $\bar{T}_1\,T_2 \Delta$ is also allowed if two different fermion triplet fields are present. }
        \label{tab:Yukawa}
        \end{center}
        \end{table}
         
        Thus scalar doublets occur if and only if we connect a fermionic doublet to a fermionic
        non-doublet. Therefore, in order to start and end up with a fermion doublet we necessarily
        have an even number of scalar doublet mass insertions. \\
        Secondly, note that in any theory including upto $SU(2)$ triplets there are only potential
        couplings possible that involve an even number of $SU(2)$ doublets. Thus, each doublet
        line will couple to an odd number of doublet lines. As the product of an even and an odd
        number is an even number,  the number of doublet lines will remain even. Connecting some
        of these lines and producing a loop will not change this fact as this closing reduces
        the number of external doublet lines by an even number. \\
        On the other hand two fundamental building rules for conformally invariant neutrino mass
        generation require that firstly there is always an odd number of mass insertions and secondly
        potential couplings always connect four lines. Both together yield that there has to be left
        an odd number of scalar external lines.
        Consequently as there has to be an odd number of VEVs but an even number of doublet VEVs,
        there has to be a singlet or a triplet VEV. Note, however, that this proof is based on
        the assumption that there are no fermion or gauge boson loops involved. This finding can be
        summarized as follows:
	If there are no gauge boson or fermion loops
        possible, a conformally invariant theory with upto $SU(2)$ triplet scalars and fermions
        needs a singlet or triplet scalar vacuum expectation value to generate left-handed
        Majorana neutrino masses.

      \subsection{ Radiative Models}
      \label{sec:Radiative}
        In this subsection we deal with the question if it is possible to choose the particle content
        and the VEV structure of a theory such that the lowest order contribution to the left-handed
        Majorana masses is fully radiative i.e. there is no scalar with a VEV coupled to the neutrino line. \\
        We assume that there are no fermion or gauge boson loops. In this case, if  in the potential  only terms appear which couple fields in singlet pairs neutrinos can not gain mass via loops. This is the case, as scalars connected to the fermion line can only be 
        coupled in such a way that they either produce one scalar of the own kind and two of another
        or couple to a particle of the own kind coming from the fermion line and thus reducing
        the number of its species by an even number. So either the number of a species stays the
        same, reduces or increases by an even number. As there has to be an odd number of mass insertions to 
        the fermion
        line it is thus impossible to combine all scalars connected to the fermion line in a loop
        without producing at least one external line that already couples to the fermion line. \\
An other way to understand this, is that for a loop induced active neutrino mass there has to be a lepton 
number violating term in the potential. Since the potential contains only four scalar operators, there has 
to be at least one among them with non pairwise coupled scalars. We can summarize this result: 
In a conformally invariant theory without fermion or gauge boson loops it is impossible
to generate left-handed Majorana neutrino masses in a fully radiative way if the potential
contains only terms coupling scalars in singlet pairs.

        We present models, which have not only pairwise scalar combinations in the potential,  and yield fully radiative left-handed neutrino masses in  \app \ref{app:Exceptions}. We only discuss models, which can yield neutrino masses with non vanishing diagonal elements, as those are excluded experimentally, as argued in \cite{Law:2013dya}. Furthermore, two possibilities to circumvent the above argument are presented, one is a model containing fermion loops.
The other is the Ma model \cite{Ma:2006km} with a $Z_2$ symmetry, which forbids the Dirac tree level coupling and violates lepton number with the sterile neutrino Yukawa term. However, we do not consider discrete symmetries in the main body of the articles and the only way to have a model with this topology is with a hidden sector symmetry. The requirement of electrically neutral VEVs makes this model only viable for generating loop induced masses for the sterile neutrinos. This possibility will be discussed later on.

\section{\label{sec:Models} Overview of viable models}     
    
    In this section we will give a summary of models in which it is possible to have neutrino masses and radiative scale symmetry breaking (RSSB). 
    The criteria for the generation of neutrino masses are presented in \sect \ref{sec:Rules}.   As we will see the RSSB works in the case that at least two additional bosonic degrees of freedom are present, of which at least one must be a scalar. This modifies the beta function of the mass parameter in such a way that a scalar component gets a VEV, which is then cascaded to the Higgs sector through the Higgs-scalar mixing as described in \sect \ref{sec:Introduction}. The symmetry breaking is consistently studied in the Gildener-Weinber approach \cite{Gildener:1976ih}, which relies on the existence of a flat direction in the classical potential. Then the one loop effictive potential is computed.  The minimal requirement of two bosonic degrees of freedom in the additional sector is crucial, since the RSSB relies on the bosonic degrees of freedom dominating over the top quark contributions. 

The symmetry breaking has to be triggered by the hidden sector and the pseudo-glodstone boson (PGB) associated with the scale symmetry breaking has to reside mainly in the hidden sector, see for example \cite{Radovcic:2014rea}. In the case of one additional bosonic degree of freedom, the Higgs boson is mainly the PGB which phenomenologically requires larger values of quartic couplings and leads to low scale Landau poles, see for example discussion in \cite{Foot:2007ay}, which corresponds to model \textbf{3A} with only one real scalar field. It was demonstrated that RSSB is possible, but in our opinion the low scale Landau pole is problematic and we will take the model with two real scalars as the simplest realistic model.

We will demonstrate the RSSB in a case with two bosonic degrees of freedom in the HS. The scalar field content is given by the $SU(2)$ doublet $H$ and two real SM singlets $\Phi$ and $S$. The potential has the form 

\begin{align}
V(H, \Phi, S) = \frac{\l_H}{2}\,(H^\dagger\,H)^2 + \frac{\l_S}{2}\,S^4 + \frac{\l_\Phi}{2} \,\Phi^4 + \\ \nonumber      \l_{HS}\,H^\dagger\,H\,S^2 + \l_{H\Phi}\,H^\dagger\,H\,\Phi^2 +\l_{S\Phi}\,\Phi^2\,S^2\,.
\end{align}

For simplicity we will use spherical coordinates in field space with the replacements

\begin{align*}
\label{eqn:fieldDefs}
H = r\, \sin \theta \sin \omega \, , \\ \nonumber
S = r \, \sin \theta \cos \omega\, ,  \\ \nonumber
\Phi = r \, \cos \theta\,.
\end{align*}

We find with \eqn \ref{eqn:fieldDefs} and the definitions $(\tan \theta)^2 =: \epsilon$ and $(\sin \omega)^2 =:\delta$ that 

\begin{align}
& (r\,\cos \theta )^4 \, V(r, \theta\, \Phi) = \frac{1}{2} \, \left( \l_\Phi + \epsilon (2\,\delta \l_{H\,\Phi} + 2 (1- \delta) \l_{S\Phi} + \right. \\  \nonumber
& \left. \epsilon (\delta^2 \l_H+ 2 (1- \delta )\,\delta \,\l_{HS} + (1-\delta)^2 \,\l_S ) ) \right)= R(\Lambda)\,.
\end{align}
    
The vanishig of this quantity at the scale of symetry breaking $R(\Lambda_{RSSB})=0$ defines the classically flat direction in the potential, it is the renormalization condition. 
  
Assuming that the mixing anomg the scalars is not large i.e. $\epsilon, \, \delta < 1$ a hierarchical VEV structure appears

\begin{align}
\ev{\Phi} & = \ev{r}  (1+\epsilon)^{-1/2} = : v \\ \nonumber
\ev{S} & = \ev{r}  (1+\epsilon)^{-1/2}  \sqrt{\epsilon} = v \, \sqrt{\epsilon}\\ \nonumber
\ev{H} &  = \ev{r} \, (1+\epsilon)^{-1/2} \sqrt{\epsilon\, \delta} = v\,\sqrt{\epsilon\,\delta} \\ \nonumber
& \Rightarrow \ev{\Phi} > \ev{S} > \ev{H} \,.
\end{align}  

After the symmetry breaking the right handed neutrinos get their Majorana mass through Yukawa interactions with the HS scalars $M_{N_{i}}= Y_{N_{i}}/2 \,v^2 (1 +\epsilon)$. The scalar spectrum contains two massive excitations and one which is mass-less on tree level and corresponds to the flat direction in the potential. The idea behind the Gildener Weinberg approach is that the quantum effects are taken into account in the one loop correction to the mass of this particle, making it a PGB of broken scale symmetry. This procedure ensures perturbativity as discussed in detail in, \cite{Gildener:1976ih}. The mass of the PGB is given by 

\begin{align}
M_S^2 = \frac{1}{8 \pi^2 \ev{r}^2}\left( M_H^4 + 6 m_W^4 + 3 m_Z^4 \right.\\ \nonumber 
\left. +M_\Phi^4 -12 m_t^4 - 2 \sum_i M_{N_{i}}^4 \right) \,,
\end{align}

while the tree level scalar masses are (with $\lambda_{\Phi S}<0$ and $\lambda_{HS}, \,\lambda_{\Phi H}>0$ for explicitness)

\begin{align}
& M_H^2  = v^2 \left[ (\delta-1 )(1 + 16 \delta \, \epsilon) \,\lambda_{\Phi\,S} + \right.\\ \nonumber
 & \left. \delta\,\epsilon (3 \delta \,\lambda_H - (\delta -1 )\lambda_{HS} )   \right]\,\delta^{-1}\,, \\
& M_{\Phi}^2  = -v^2 \left[  (16 (\delta-1 )\epsilon -1)\lambda_{\Phi S} \right.\\ \nonumber
& \left.-\epsilon (\delta \lambda_{HS} -3 (\delta-1 )\lambda_S) \right] \,.
\end{align}

\begin{figure}[h]

  \includegraphics[width=0.45\textwidth]{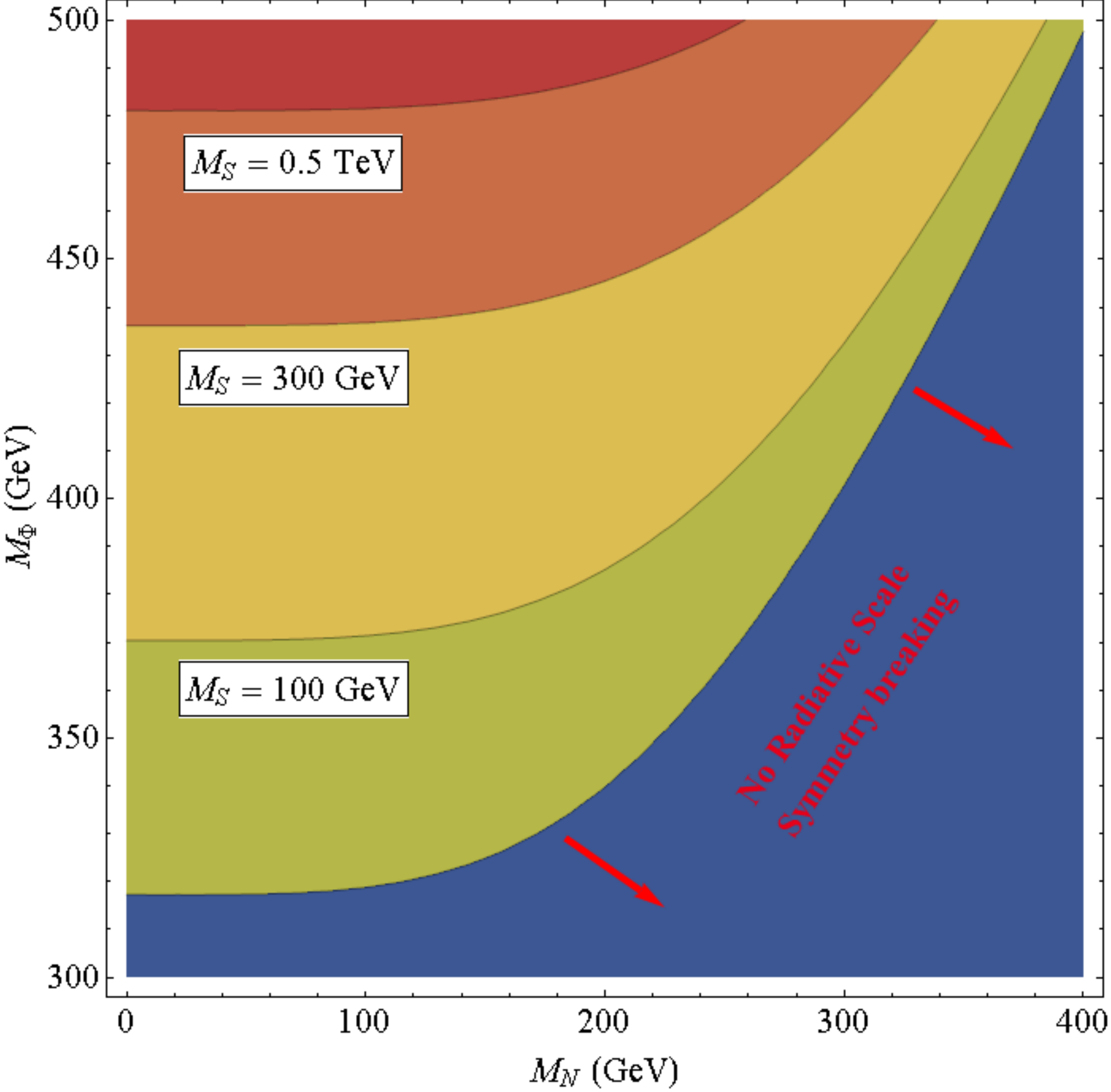}
  \caption{The phenomenlogically allowed mass region in the simplest neutrino mass model with RSSB, a Higgs mass of $125$ GeV, a higgs portal mixing compatible with the bound $\sin \theta < 0.37$, perturbative potential parameters and no low scale Landau pole. Here $M_N$ is the mass of the heaviest right handed neutrino, $M_\Phi$ is the heavy scalar dominating the spectrum and $M_S$ is the mass of the PGB.} 
  \label{fig:ThreeMasses}
\end{figure}

As can be seen the PGB resides mainly in the HS and thus the mixing with the Higgs can be brought in agreement with the experimentally constrained Higgs-scalar mixing \cite{Farzinnia:2014xia}, while the potential parameters are perturbative and no low energy Landau pole appears. We plot the phenomenologically allowed mass regions in \fig \ref{fig:ThreeMasses}.  

   Next we study neutrino mass models with RSSB, which will be organized in the following way. Firstly we distinguish models with the SM model gauge group and secondly those where an additional hidden sector symmetry comprises with the SM symmetry group a direct product group. In the first case models are distinguished which affect the left handed neutrino mass
    directly (\textbf{$\#$A}) and those with an additional singlet fermion state which contributes to the left handed neutrino masses, as known from the type I see-saw
    mechanism (\textbf{$\#$B}). 
    
    In the second scenario in all models there are additional SM singlet fermion states. We distinguish models with effect on the masses 
    of the total singlets under the full gauge group (\textbf{$\#$C}), denoted by $\nu_R$ and those where masses of fermions are affected, which carry a hidden sector
    charge (\textbf{$\#$D}) and are denoted by $\nu_x$.
    The Dirac type masses in our framework are always determined by Yukawa couplings $y_D$ and the Higgs VEV, 
    and assumed to exist if allowed by the symmetry. We comment on loop effects in models where those can lead to suppression of mass matrix entries. 	
    Furthermore, some comments on phenomenological implications will be made, but the main phenomenological discussion is omitted at this point and postponed 
    to \sect \ref{sec:Phenomenology}. All models carry an identification number and are described in detail in \app \ref{app:Models}. 
    
    At first we focus on the models where the SM field content is extended. Assuming that we have singlet, doublet and triplet fermionic and scalar $SU(2)$   representations we list all combinations systematically and check whether a conformal neutrino mass model can be constructed, see Table \ref{tab:ModelsSM}. Assuming only the above mentioned representations the presented list is complete. The models share features with the non-conformal analogues, nevertheless the scalar sector is in all cases enlarged to make the graph construction topologically possible without explicit mass insertions. Furthermore, the mass scales are all around the TeV scale, since the general spirit of the radiativly broken scale invariance forbids large scale separation.
    
We again present a full catalogue of models with a $U(1)_\text{hidden}$, given that we only involve up to the triplet representation of the $SU(2)_L$ group, see Table \ref{tab:ModelsHS}. This model sector could be enlarged by regarding more complex Hidden groups, but due to our little knowledge of the dark sector we stick here to the minimality condition. As a result we find a variety of tree level and radiative models with possible textures in the neutrino mass matrix. As one of the most promising models we point out \textbf{1D} and \textbf{2D}, which lead to an inverse see-saw (ISS) mass matrix structure which implies seizable active sterile mixing, discussed in \cite{Deppisch:2004fa, Abada:2014vea}. The active-sterile mixing and the light masses are given by  

\begin{align}
\label{ISSrelations}
\epsilon = \frac{1}{2} m^{\dagger}_{D} (M^{-1}_{Rx})^*(M^{-1}_{Rx})^T m_D \approx \frac{y_D^2}{y_M^2} \frac{v^2}{\text{TeV}^2}, \\ \nonumber
m_\nu = m^T_D (M^{-1}_{Rx})^T \mu M_{Rx}^{-1} m_D  \approx \mu \,\epsilon.
\end{align}
The $M_{Rx}$ scale is of the order of one to few TeV and the $\mu$ scale is loop induced in 2D and suppressed by heavier scales in 1D, which brings it to the keV scale. The Yukawa couplings in this region can be close to one, which makes it an attractive alternative to the fine tuned solutions. The effects of the active-sterile mixing can lead to an improved $\chi^2$ for the Electro-weak precision observables, as shown in \citep{Akhmedov:2013hec} and we will comment on it in the next section. 

In general the requirement of no scalar scale hierarchy restricts the vacuum expectation values of the new scalars not to be higher than the TeV scale. This leads with Yukawa couplings in the perturbative region to a particle spectrum below the TeV scale. However, this is not a necessity in all models. For instance if several additional scalar VEVs induce a cascade where the heaviest field begins with the symmetry breaking and transfers the scale by a portal to the next which in turn cascades down to the third scale, the scale separation can become larger without large tuning of the couplings, this can lift up the spectrum to a few TeV, as can be the case in the conformal inverse see-saw.  

In several models, see Table \ref{tab:ModelsSM} and \ref{tab:ModelsHS},  the Majorana contribution to the light neutrino mass is suppressed and therefore an other neutrino mass scenario appears, in that case the active neutrinos are almost mass degenerate with the sterile components comprising pseudo Dirac pairs. This possibility is experimentally extremely challenging, but might be accessible in long baseline and low energy oscillation experiments \cite{Beacom:2003eu}. 
 
In general scale separation does not appear naturally in models with RSSB, thus the neutrino mass scale can appear if the Yukawa couplings are arranged in a way leading to a see-saw suppression. The other possibility is that the lightness is connected to a small lepton number violation parameter. This smallness can be argued to be natural in t'Hoft sense, as the symmetry of the theory would increase if this parameter would be exactly zero. Furthermore, in radiative neutrino mass models the smallness of the lepton number violation is augmented by a mass suppression by the loop factors. The most interesting possibility is, however, if both of this mechanisms are at work. This is the case if the Majorana scale is induced by a loop involving a lepton number violating coupling, leading to the Pseudo Dirac and Inverse see-saw scenarios. Where in the last scenario the Yukawa couplings can be of order one.

\begin{widetext}
    
  \begin{table}

   \centering
  \Large{\textbf{Conformal Mass Models within the SM Gauge Group}} \normalsize
  \vspace{1mm} \\ 
  \hspace*{-2cm}

  \begin{tabular}{|>{\centering}m{1cm}|>{\centering}m{3.5cm}|>{\centering}m{3cm}|>{\centering}m{1.5cm}
    |>{\centering}m{1.8cm}|>{\centering \arraybackslash}m{6cm}|}
     \hline
   \# & particle content & non-conformal motivation & neutrino masses & correct Higgs mass &
    phenomenological note \\
    \hline
  \end{tabular}
  \vspace{1mm} \\
    \underline{\large{\textbf{Left-Handed Majorana Masses}}} \normalsize \\
    \centering
    \vspace{1mm} 
  \begin{tabular}{|>{\centering}m{1cm}|>{\centering}m{3.5cm}|>{\centering}m{3cm}|>{\centering}m{1.5cm}
    |>{\centering}m{1.8cm}|>{\centering \arraybackslash}m{6cm}|}
    \hline
    1A & Conformal SM (CSM) & $\diagup$ & 
    \textbf{No} & \textbf{No} & This theory does not yield neutrino masses. \\
    \hline
    2A & CSM + $\nu_R:(1,0)$ & See-saw type I & \textbf{Yes} & \textbf{No}
    & Neutrinos in
    this theory are of Dirac type. \\
    \hline 
    3A & CSM + $\nu_R:(1,0)$ + $\varphi:(1,0)$ & See-saw type I & \textbf{Yes} 
    &    \textbf{Yes}
    & In dependence of the coupling constants this theory can yield Sub TeV or Pseudo-Dirac 
    neutrinos. \\
    \hline
    4A & CSM + $\Delta:(3,-2)$ & See-saw type II &  \textbf{Yes} & \textbf{No}
    & This theory yields pure left-handed Majorana neutrinos.\\
    \hline 
    5A & CSM + $\Delta:(3,-2)$ + $\varphi:(1,0)$ & See-saw type II &  \textbf{Yes} 
    &  \textbf{Yes} & This theory yields pure left-handed Majorana neutrinos as well. \\
    \hline
    6A & CSM + $\nu_R:(1,0)$ + $\varphi:(1,0)$ + $\Delta:(3,-2)$ & See-saw type I/II
    &  \textbf{Yes} &  \textbf{Yes} & Sub TeV and Pseudo-Dirac neutrinos are
    possible. \\
    \hline
    7A & CSM + $\delta_-:(1,-2)$ & $\diagup$ & \textbf{No} & \textbf{No}
    & Neutrinos remain massless. \\
    \hline
    8A & CSM + $\delta_-:(1,-2)$ + $\Delta:(3,-2)$ & $\diagup$ &  \textbf{Yes} & \textbf{No}
    & The additional $\delta_-$ only contributes corrections to the masses. \\
    \hline
    9A & CSM + $\Sigma:(3,0)$ & See-saw type III & \textbf{No} & \textbf{No}
    & Neutrinos remain massless. \\
    \hline
    10A & CSM + $\Sigma:(3,0)$ + $\varphi:(1,0)$ & See-saw type III &  \textbf{Yes} & 
     \textbf{Yes} & This theory yields the same neutrino phenomenology like the conformal
    See-saw type I. \\
    \hline
    11A & CSM + $\delta_-:(1,-2)$ + $\epsilon_{++}:(1,4)$ + $\varphi:(1,0)$ & Zee-Babu &
     \textbf{Yes} &  \textbf{Yes} & Pure left-handed Majorana neutrino masses 
    suppressed by 2 loops. \\
    \hline
     12A & CSM + $H_2:(2,1)$ + $\eta_{+}:(1,2)$ + $\varphi:(1,0)$ & Zee Model &
     \textbf{Yes} &  \textbf{Yes} & Pure left-handed Majorana neutrino masses 
    suppressed by 1 loop. \\
    \hline
     13A & CSM + $\phi_1:(2,3)$ $ H_2:(2,1) $ $ \eta:(1,-4) ;\; \phi_2:(1,0)$ &Law-McDonald &
     \textbf{Yes} &  \textbf{Yes} & Pure left-handed Majorana neutrino masses 
    suppressed by 2 loops. \\
    \hline
    \end{tabular}

    \vspace{1mm} 
    
    \underline{\large{\textbf{Right-Handed Majorana Masses}}} \normalsize \\
    \vspace{1mm} 
    \begin{tabular}{|>{\centering}m{1cm}|>{\centering}m{3.5cm}|>{\centering}m{3cm}|>{\centering}m{1.5cm}
    |>{\centering}m{1.8cm}|>{\centering \arraybackslash}m{6cm}|}
    \hline
    1B & CSM + $\nu_R:(1,0)$ + $\Sigma:(3,0)$ + $\Delta:(3,0)$ + $\varphi:(1,0)$ & $\diagup$ &
     \textbf{Yes} &
     \textbf{Yes} & This theory can generate conditions for the Pseudo-Dirac and the Sub TeV see-saw. \\
    \hline 
    2B & CSM + $\nu_R:(1,0)$ + $\nu_x:(1,0)$ + $\varphi:(1,0)$ & $\diagup$ &  \textbf{Yes} 
    &  \textbf{Yes} & The extension by further sterile neutrinos is trivial if they cannot
    be distinguished from the original sterile neutrinos. \\
    \hline
  \end{tabular}
  \caption{
  \label{tab:ModelsSM}  
  Summary of different conformally invariant models for the generation of neutrino masses
  within the SM gauge group. The Lorentz nature of the fields is the following: $\nu_R,\,\Sigma$  are fermions and $\phi,\,\delta,\,\epsilon,\,\Delta$ 
  are scalars. It is always mentioned if there is a non conformal motivation to the particular model. Furthermore, short comments
  on the phenomenology are displayed. All models carry a number for later reference. \\
Note that in the case of model \textbf{3A} with one real singlet scalar RSSB is possible, however the theory has a low scale Landau pole. Thus, we consider the model to be only phenomenologically viable if at least two real scalars are present, as discussed in \sect \ref{sec:Models}.}
 
\end{table}

\begin{table}
  \Large{\textbf{Conformal Mass Models with Additional U(1) Symmetry}} \normalsize 
  \vspace{1mm} \\
  \begin{tabular}{|>{\centering}m{1cm}|>{\centering}m{3.5cm}|>{\centering}m{1.5cm}|
  >{\centering}m{3.5cm}|>{\centering\arraybackslash}m{5cm}|}
    \hline
    \# & particle content & $U(1)_H$ & VEV structure & phenomenological note \\
    \hline
  \end{tabular}
  \vspace{1mm} \\
    \underline{\large{$\boldsymbol{\nu_R}$ \textbf{Majorana Masses}}} \normalsize \\
    \vspace{1mm} 
  \begin{tabular}{|>{\centering}m{1cm}|>{\centering}m{3.5cm}|>{\centering}m{1.5cm}|
  >{\centering}m{3.5cm}|>{\centering\arraybackslash}m{5cm}|}
    \hline
    & $\nu_R:(1,0)$ & 0 & & The double see-saw mass structure is implied. \\
    1C & $\nu_x:(1,0)$ & 1 & all scalars get a VEV & Pseudo-Dirac  and \\
    & $\varphi_1:(1,0)$ & 1 & &  sub TeV scenarios \\
    & $\varphi_2:(1,0)$ & 2 & &   are possible . \\
    \hline
    & $\nu_R:(1,0)$ & 0 & & The minimal extended see-saw structure is implied. \\
    2C & $\nu_x:(1,0)$ & 2 & all scalars get a VEV & Light sterile neutrinos \\
    & $\varphi_1:(1,0)$ & 0 & &  with large \\
    & $\varphi_2:(1,0)$ & -2 & &   active-sterile mixing . \\
    \hline
    3C & theory 1C + & theory 1C & $\varphi_1$ gets no VEV & radiative model, \\
    & $\varphi_3:(1,0)$ & -4 & & implies Pseudo-Dirac scenario \\
    \hline
    & $\nu_R:(1,0)$ & 0 & &  \\
    & $\Sigma:(3,0)$ & 1 &  & Pseudo-Dirac and sub TeV\\
    4C & $\Delta:(3,0)$ & 1 & all scalars get a VEV & scenarios \\
    & $\varphi_1:(1,0)$ & 1 & & are possible. \\
    & $\varphi_2:(1,0)$ & 2 & & \\
    \hline
    5C & theory 3C + & theory 3C & $\varphi_1$ gets no VEV & radiative model, \\
    & $\varphi_3:(1,0)$ & -4 & & implies Pseudo-Dirac scenario \\
    \hline
   \end{tabular}
   
   \vspace{1mm} 
    \underline{\large{$\boldsymbol{\nu_x}$ \textbf{Majorana Masses}}} \normalsize \\
   \vspace{1mm} 
  \begin{tabular}{|>{\centering}m{1cm}|>{\centering}m{3.5cm}|>{\centering}m{1.5cm}|
  >{\centering}m{3.5cm}|>{\centering\arraybackslash}m{5cm}|}
    \hline
    & $\nu_R:(1,0)$ & 0 & & \\
    & $\nu_x:(1,0)$ & 1 & & \\
    1D & $\Sigma:(3,0)$ & -2 & all scalars get a VEV & generates small $\nu_x$ mass, \\
    & $\Delta:(3,0)$ & -3 & & implies the inverse see-saw scenario \\
    & $\varphi_1:(1,0)$ & -3 & & \\
    & $\varphi_2:(1,0)$ & -4 & & \\
    & $\varphi_4:(1,0)$ & 1 & & \\
    \hline
    2D & theory 1D + & theory 1D & $\varphi_1$ gets no VEV & radiative model, \\
    & $\varphi_3:(1,0)$ & 10 & & implies the inverse see-saw scenario \\
    \hline
  \end{tabular}
  \caption{
 \label{tab:ModelsHS}  
  Summary of different conformally invariant models for the generation of neutrino masses
  with an additional HS symmetry.  The Lorentz nature of the fields is the following: $\nu_R,\,\nu_x,\,\Sigma$  are fermions and $\phi_i$ for 
  $i \in \{1,2,3\}$ are scalars. All hidden sector charges are  shown in the third column and the VEV structure is summarized in the fourth column.
  Furthermore, short comments on the phenomenology are displayed. All models carry a number for later reference.}
\end{table}

\end{widetext}

\section{\label{sec:Phenomenology}Phenomenology}

In this section we will check which of the proposed models can indeed reproduce the correct neutrino mass phenomenology i.e. the mass square differences and the correct mixing angles and at the same time be consistent with rare decay experiments and electroweak precision observables (EWPO). In a plot we will demonstrate viable regions of the allowed parameter space and estimate expected signals for future lepton flavour and number violation experiments. 

In most of the discussed models the PMNS matrix becomes not exactly unitary. This happens if the active-sterile mixing is considerable and induces a number of effects on physical quantities as the Weinberg angle, the W-boson mass, the left and right handed couplings $g_L$, $g_R$, the leptonic and invisible Z-boson decay width and the neutrino oscillation probabilities, for more detailed discussion and limits see \cite{Antusch:2006vwa} and references therein. Thus studying the non unitarity allows to narrow down the parameter space of a given model. However, some effects are not captured by this treatment only. Those are processes where explicit particle propagation is responsible for the new physics signal. To get an order of magnitude estimate we integrate out heavier degrees of freedom to obtain an effective scenario with a $(3+n)\times(3+n)$ nearly unitary mixing matrix $\mathbf{U}$. 
 
 \begin{equation}
    \label{mixingmatrix}
    \mathbf{U}=\left(
    \begin{array}{cc}
    \mathscr{U} & \mathscr{R}\\
    \mathscr{W} & \mathscr{V}
    \end{array}
    \right)\,.
  \end{equation}
  
This corresponds to a scenario with three active and $n$ sterile neutrinos. Here $\mathscr{R}$ can be considered as the active-sterile mixing. $\mathscr{U}$ is the PMNS matrix and is not unitary any more. A measure for non-unitarity of the PMNS matrix  is given by
  \begin{align}
    \label{epsilon}
    \e_\alpha\equiv {\textstyle\sum_{i\geq 4}} |\mathbf{U}_{\alpha i}|^2\,.
  \end{align} 

This matrix diagonalizes the following mass matrix

 \begin{equation}
    \mathcal{M}=\begin{pmatrix} m_L & m_D\\ m_D^T & M_R\end{pmatrix}.
  \end{equation}
  
Thus, the active and sterile neutrinos have a Majorana mass and mix due to the Dirac mass terms. This set up covers all the effects on neutrino physics of a given model. The Majorana mass nature opens the possibility for lepton number violation and neutrino-less double beta decay. The propagating sterile states lift the GIM suppression in the lepton flavour violating processes for the charged leptons and different non-unitraity parameters $\epsilon_\alpha$ in \eqn \ref{epsilon} parametrise deviation from lepton universality. Furthermore, in this set up we can get estimates on the oblique corrections \cite{Peskin:1990zt}. As shown in \cite{Akhmedov:2013hec} they can contribute significantly to EWPOs especially given large non-unitarity and heavy sterile neutrinos.

The mass terms in the effective theory after integrating out heavier degrees of freedom have the following form.  
  \begin{align}
    \label{eq:MassTerms}
    -\mathscr{L}_m =\frac{1}{2} m^{*}_{\!L,ij} \bar{\nu}^{c}_{\!L\!,i}\nu^{}_{\!L\!,j} + m^{*}_{\!D,ij}\bar{\nu}^{}_{\!L,i}\nu^{}_{\!R,j} \\ \nonumber
    +\frac{1}{2}M^{*}_{\!R,ij}\bar{\nu}^{c}_{\!R\!,i}\nu^{}_{\!R\!,j}+h.c.,
  \end{align} 
where $m^{*}_{\!D,ij}=g^{}_{H\!,ij}\cdot v_H$  and $M^{*}_{\!R,ij}=g^{}_{\varphi\!,ij}\cdot v_\varphi$. While direct masses for the left handed neutrinos are generated due to
 a scalar or fermionic triplet.  $m^{*}_{\!L,ij}=g^{a}_{L\!,ij}\cdot v_\Delta  + (g^{}_{\Sigma\!,ij}+g^{b}_{\Delta\!,ij})\cdot v_H^2/v_\phi$.  The scalar triplet contributes a dimension four operator, both triplets generate terms proportional to the squared Higgs VEV, the fermionic triplet a dimension fife operator  and the scalar triplet a dimension six operator. The $g$ parameters are effective Yukawa couplings which can contain corrections from heavier particles integrated out of the theory. Thus depending on the theory in question the perturbativity condition is not to be taken as a strict bound. 
  
  Scanning over the effective Yukawa coupling space provides us with sets of viable solutions according to the above criteria. To visualize the solutions we set up a two dimensional map with the horizontal axis for the averaged right handed mass scale and the vertical axis for the averaged Dirac couplings. The average represents the order of magnitude of the Yukawas in the case they are in the same ball park, if they are spread apart the average is dominated by the largest. The spread over several orders of magnitude, however, is not considered as it would require unnaturally large tuning. The results of our study are presented in \fig \ref{fig:Masterplot}.  We would like to discuss four phenomenological scenarios separately.

\subsection{Pure left handed Majorana mass}
In the case that the Dirac coupling is very small, or there are no fermionic singlets under the SM gauge group included in the theory, the only possible neutrino mass term is the left handed Majorana mass. In this scenario the charged lepton flavour violation is strongly GIM suppressed and beyond experimental precision. The PMNS mixing matrix is unitary and therefore the most promising signals are expected in the $0\nu\beta\beta$ experiments. The total mass scale enters the effective electron neutrino mass, since it is entirely Majorana. For a detailed study if the $0\nu\beta\beta$ sensitivity depending of the hierarchy see \cite{Rodejohann:2012xd} and references therein. The current experimental bound on the electron neutrino effective mass, the parameter controlling the $0\nu\beta\beta$ decay, is $\ev{m_{ee}}< 0.4 \, \text{eV}$  \cite{Macolino:2013ifa}. 

The models leading to a pure left handed Majorana mass are in our set up \textbf{11A}, \textbf{12A} and \textbf{13A}  here the neutrino masses are suppressed by one or two loops.

\subsection{Pseudo Dirac Scenario}

The other distinct region with only light neutrinos is around the point in the parameter space where the active and sterile neutrinos form mass degenerate pairs of Dirac fermions. This point has no lepton number violation and an effective GIM suppression of the charged lepton flavour violating process. Now the pairs can acquire a small Majorana mass, either through contributions of light sterile neutrinos or a small $m_L$ mass.

This leads to a mass splitting among the degenerate Majorana pairs forming the effective Dirac neutrino states. Phenomenologically this is consistent with observations as long as this splitting is smaller than the experimental accuracy of the mass square difference measurement. For detailed bounds consult \cite{deGouvea:2009fp}. It turns out that the strongest constraints apply to the splitting of the first and second mass state and are of the order of $10^{-9}$ eV, while in the third mass state, with the dominant tau flavour, the splitting can be up to $10^{-3}$ eV.  Since the right-handed neutrinos are light in this scenario, the PMNS matrix is unitary and there are no phenomenological bounds from EWPOs, lepton universality or lepton flavour violation.

The origin of the mass splitting is not important for oscillation experiments, when it comes to lepton flavour violation, however, there is an interesting subtlety. Consider the two cases in the one flavour scenario, where in the first case the Majorana mass appears on tree level for the active neutrino and in the second case for the sterile component

 \begin{equation}
    \mathcal{M}_1=\begin{pmatrix} \mu & m_D\\ m_D & 0\end{pmatrix}  \text{  and   }  \mathcal{M}_2=\begin{pmatrix} 0 & m_D\\ m_D & \mu\end{pmatrix}.
  \end{equation}
  
In the limit $\mu << m_D$ the mass eigenvalues are given in both cases as $m_\pm = \pm m_D +\mu/2 $ and the diagonalizing mixing matrices are

 \begin{equation}
    U_{1/2} \approx \sqrt{\frac{1}{2}} \begin{pmatrix} 1 \pm \epsilon & -1 + \epsilon \\ 1 \mp \epsilon & 1 + \epsilon \end{pmatrix}  \text{  with   } \epsilon = \frac{\mu}{4 m_D}.
  \end{equation}
  
  \FloatBarrier
  
We consider now the expansion of the effective mass for the neutrnoless double beta decay in powers of the momentum transfer. The effective mass is approximately given by  $\ev{m_{ee}} \approx |q^2 \textstyle{\sum_{i}} \mathbf{U}_{e i}^2 \, m_i/(q^2 - m_i^2)|$, the relation holds that $ m_i^2 \ll  |q^2| \approx  (0.1 \,\text{GeV})^2$ and we can expand 

\begin{align}
\ev{m_{ee}} \approx |\sum_i U_{ei}^2 m_i + 1/q^2 \sum_i U_{ei}^2 m_i^3 + O(1/q^4)|.
\end{align}

Inserting the parameters we find that in the case where the active neutrino has a direct Majorana mass the effect is of order \cite{Rodejohann:2012xd},
\begin{align}
\label{eqn:directCont0NuBetaBeta}
\ev{m_{ee\, 1}} \approx \mu \approx (m_+^2 - m_-^2)/(2 \, m_D) \,.
\end{align}

 In the case where the Majorana mass appears via the sterile component, the first order contribution vanishes, as the electron neutrino entry of the neutrino mass matrix is zero and one has to leading order
 \begin{align}
 \ev{m_{ee\,2}} \approx m_D\, (m_+^2 - m_-^2)/(2 \, q^2)  \approx \mu \, m_D^2/q^2 \,,
 \end{align}
since $m_D$ is of the order of the absolute neutrino mass scale the effective mass is suppressed by the factor $(m_D/q)^2$ with respect to  \eqn \ref{eqn:directCont0NuBetaBeta} which is at least 14 orders of magnitude. Thus there is an interesting experimental possibility to distinguish these scenarios. Assume, neutrino oscillation experiments on cosmic scales detect a small mass splitting testing oscillations on cosmic length scales as described in \cite{Beacom:2003eu}. If this splitting is in the phenomenologically allowed region today, the contribution to the effective mass for $0\nu\beta\beta$ decay can be up to a few $10^{-5}$ eV, as displayed in   \fig \ref{fig:Masterplot} in the zoomed in region. This is only the case, if the mass splitting originates from a direct active neutrino Majorana mass, in the second case it would be of the order $10^{-17}$ eV and beyond experimental reach. Thus, measuring the $0\nu \beta \beta$ decay provides evidence of scenario one and placing a limit smaller than the predicted value shows that scenario two is realized. Even though the possibility is interesting from the theoretical perspective, it is extremely challenging experimentally, since the maximal expected decay rate is four orders of magnitude below the current sensitivity.

Among the presented models the pseudo Dirac scenario can be realized in \textbf{3A}, \textbf{6A} with a direct active neutrino Majorana mass and in 1B and possibly in all the (\textbf{$\#$C}) models with a sterile neutrino Majorana mass, leading to no observable $0\nu\beta\beta$ decays. The Yukawa couplings in the Majorana and Dirac sector have to be tiny, in fact below $10^{-12}$, but there is no hard theoretical argument which could exclude this possibility a priori.

\subsection{Sub-TeV Yukawa See-saw}

In several models where the right handed Majorana couplings are not loop or scale suppressed the right handed Majorana mass is below one TeV. Viable solutions lie in a triangular shaped region for Majorana couplings smaller than one. In this parameter region the contributions to $m_L$ are subdominant and thus the averaged effective Yukawa couplings represent the sterile Majorana mass on the x-axis and the Dirac mass on the y-axis respectively. For the scan the Casas-Ibarra parametrisation  \cite{Ibarra:2010xw} was used, which parametrizes the active sterile-mixing as $R = -i D_{\sqrt{m_\nu}}\mathcal{O}^*D_{\sqrt{M_R}}U_{\text{PMNS}}$ with $\mathcal{O}^T\,\mathcal{O}=1$. Thus the physical effects connected to non-unitarity are controlled by the norm of $\mathcal{O}$.

The shape of the Sub-TeV see-saw region is explained as follows. The see-saw relation ($m_D^2/M_R \approx m_\nu = 0.1$ eV) and $M_R > m_D$ sets the lowest value for $m_D$ given a $M_R$, which is  $m_D > \sqrt{m_\nu \, M_R}$. This sets the lower boundary of this region and is represented by a black dotted diagonal line of gradient two in the log plot in \fig \ref{fig:Masterplot}. Deviations to higher Dirac couplings induce a larger active-sterile mixing and thus larger non-unitarity. The unitarity bounds constrain the region to the left and are represented by a brown line, for non-unitarity of one percent. Since deviation from unitarity is proportional to $m_D/M_R$ the line has gradient one in the log plot. However, those turn out to be not the strongest constraints. The most stringent bounds come from neutrino-less double beta decay, displayed as a red line. In the region of interest the dominant contribution to the electron neutrino effective mass for $0\nu\beta\beta$ is given by $ \ev{m_{ee}} \approx | \textstyle{\sum_{i\geq 4}} \mathbf{U}_{e i}^2 \,\text{GeV}^2/M_i|$. The predicted effective electron neutrino mass violates the observational bound if the right handed neutrinos become too light. The constraints from rare lepton flavour violating decays and lepton universality are somewhat weaker. The EWPOs in this parameter region are consistent with their measured values.  To this end the $\chi^2$ function as in \citep{Akhmedov:2013hec} has been calculated and loop effects of the right handed neutrinos included, the resulting $\chi^2$ values do not differ significantly from the SM values. 

\begin{figure}[h]

  \includegraphics[width=0.45\textwidth]{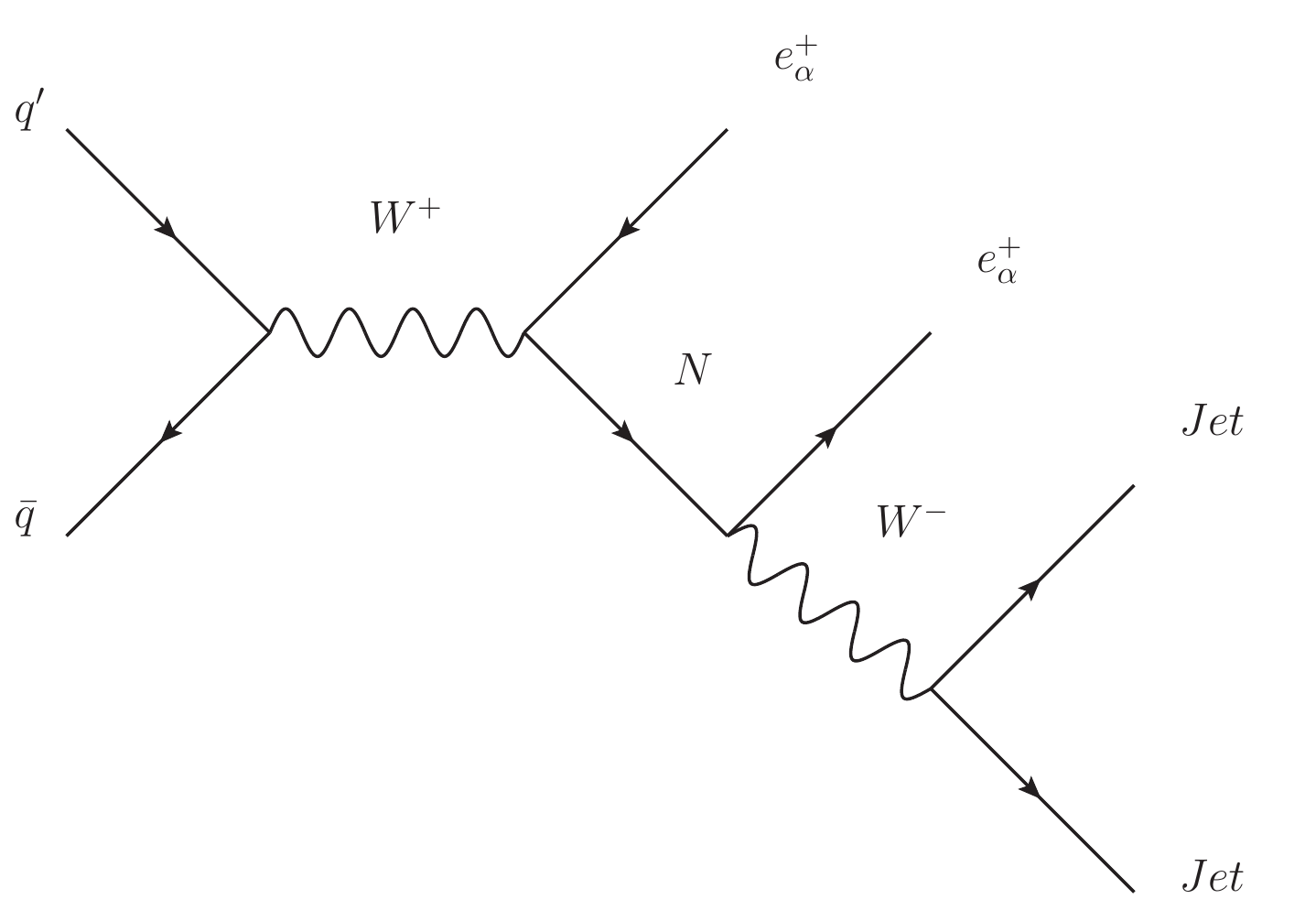}
  \caption{The lepton number violating decay as a collider signature for the Sub-TeV and multi TeV see-saw with a heavy Majorana Neutrino decay.} 
  \label{fig:DecayLNV}
\end{figure}

The main characteristics of this scenario is lepton number violation, since the Majorana mass of the sterile neutrino is not suppressed. Besides the rare decay processes, lepton number violation can lead to beyond SM processes at colliders in decays of the heavy Majorana neutrinos, see \fig \ref{fig:DecayLNV}. The production cross section for this process, is proportional to $ | \textstyle{\sum_{i}} \mathbf{U}_{\alpha i}^2 \,M_i^{-1}|$ \cite{Keung:1983uu} and has basically zero SM background.

It is interesting to discuss the values of the Yukawa couplings in this region. Since in our models of spontaneous broken scale invariance all the masses are a result of a VEV coupled with a Yukawa term, the see-saw relation is induced entirely by the Yukawa coupling structure. While for the Majorana coupling the region of the Sub-TeV see-saw implies in the presented models couplings between $10^{-3}$ and one, the Dirac Yukawa couplings vary between $10^{-7}$ and $10^{-4}$. This values might be considered small and fine-tuned, however we have to stress here that in that case the electron Yukawa coupling, which is of order $10^{-6}$ is suspicious as well. In the models discussed above this scenario is realized in \textbf{3A}, \textbf{6A}, \textbf{10A} , \textbf{1B}, \textbf{1C} and \textbf{4C}.

\subsection{Inverse Yukawa see-saw}

The most interesting scenario from the theoretical point of view in the context of RSSB is the inverse see-saw, introduced in \cite{Deppisch:2004fa, Abada:2014vea}. It naturally occurs in models \textbf{D1} and \textbf{D2}, where the mass matrix has the following texture and the scale $\mu$ is loop or scale suppressed 

 \begin{equation}
 \label{eq:seesawInv}
        \mathcal{M} = 
        \begin{pmatrix}
          0 & m_D & 0 \\
          m_D^T & 0 & M_{Rx} \\
          0 & M_{Rx}^T & \mu
        \end{pmatrix} \, .
      \end{equation}
       
The spectrum of this models contains Pseudo Dirac pairs of heavy neutrinos, with masses of order $M_{Rx}$ and their mass splitting $\mu$ determines the amount of lepton number violation present. At the same time it is the parameter, which controls the smallness of the active neutrino masses.  As given by \eqn \ref{ISSrelations} the active sterile mixing is determined by the ratio $m_D^2/M_{Rx}^2$ and the general spirit of RSSB together with no tuning in the Yukawa couplings suggests seizable  values. Seizable mixing is only compatible with small active masses if a cancellation mechanism is at work. It can be seen in the Casas-Ibarra parametrization, we choose the two flavour case with $U_\text{PMNS}=1$ for simplicity here. The orthogonal complex matrix $\mathcal{O}(\theta)$ is in this case a simple $2\times 2$ rotation matrix with the complex angle $\theta = a +ib$. In the limit $a \ll 1$ and $1 \ll b$ we have

 \begin{equation}
 \label{eq:ISSIbarra}
        \mathcal{O}(\theta) \approx 
        \begin{pmatrix}
          \cosh(b) & - i\sinh(b)  \\
          i \sinh(b) & \cosh(b) 
        \end{pmatrix} \, .
      \end{equation}
 
As shown in \citep{Kersten:2007vk}, this leads  in the limit of $m_\nu \rightarrow 0$ and with $\sinh(b)\approx \cosh(b) \approx e^b$, $e^b\sqrt{m_1} \rightarrow \sqrt{\mu}, \,\,e^b\sqrt{m_2} \rightarrow \alpha \sqrt{\mu}$ to an  

 \begin{equation}
 \label{eq:ISSIbarra}
        m_D \approx 
        \begin{pmatrix}
          \sqrt{\mu M_1} & - i\sqrt{\mu M_2}  \\
         i\sqrt{\mu M_1}\alpha & \sqrt{\mu M_2}\alpha 
        \end{pmatrix} \, .
      \end{equation}
Which in the limiting case has rank 1 and thus induces massless active neutrinos. This shows that the orthogonal matrix with dominating imaginary arguments is a good effective description of the ISS.   
 
Using this fact we study experimental constraints on this scenario. At first we consider the $0\nu\beta\beta$ decay, which placed the most severe bounds on the Sub-TeV scenario. The general expression useful to consider in this case is \cite{Abada:2014vea} $\ev{m_{ee}} \approx |q^2 \textstyle{\sum_{i}} \mathbf{U}_{e i}^2 \, m_i/(q^2 - m_i^2)     |$ . Which now can be studied in three cases, depending on the ratio of $q^2/M_{Rx}^2$, where the neutrino momentum is $|q| \approx  0.1 \,\text{GeV}$. 

If we have $M_{Rx} \gg 0.1 \text{GeV}$ and using the facts that for  $i>3$, $\mathbf{U}_{e i}^2 \approx m_D^2/M_{Rx}^2$  and $ \mu\,m_D^2/M_{Rx}^2  \approx m_\nu $ the following approximation holds
\begin{align}
\ev{m_{ee}} \approx \left|\textstyle{\sum_{i=1}^3} \mathbf{U}_{e i}^2 \, m_i   - \frac{q^2}{2}  \textstyle{\sum_{i>3}} \mathbf{U}_{e i}^2 \frac{\mu}{m_i^2} \right| \\ \nonumber
\approx  \left|\textstyle{\sum_{i=1}^3} \mathbf{U}_{e i}^2 \, m_i   - m_\nu \frac{q^2}{M_{Rx}^2} \right| \approx \left|\textstyle{\sum_{i=1}^3} \mathbf{U}_{e i}^2 \, m_i  \right|.
\end{align}
Which means that the rate is purely given by the light neutrino spectrum with well known phenomenology. 

The other limit is $M_{Rx} \ll 0.1 \,\text{GeV}$, leading to $\ev{m_{ee}} \approx | \textstyle{\sum_{i}} (\mathbf{U}_{e i}^2 \, m_i  +  1/q^2\,\mathbf{U}_{e i}^2 \, m_i^3)    | = \mathcal{M}_{ee}+O(\mu m_D^2/q^2)$. This situation is similar to the discussed Pseudo Dirac scenario with light neutrinos and the lowest order contribution is  $\mu m_D^2/q^2 < \mu M_{Rx}^2/q^2$, which in this limit is negligible. 

The only case when the heavy Pseudo Dirac states can measurably contribute to the $0\nu\beta\beta$ decay is when $M_{Rx} \approx  0.1\, \text{GeV}$. Then we have 

\begin{align}
\ev{m_{ee}} \approx \left|m_{ee}^\text{light} + \textstyle{\sum_{i>3}} \mathbf{U}_{e i}^2 \,\mu   \left( 1 + \frac{ m_i^2}{|q^2|} \right)^{-1} \right| \\ \nonumber
\approx   \left| m_{ee}^\text{light} + \textstyle{\sum_{i>3}}  m_\nu \, \left( 1 + \frac{ m_i^2}{|q^2|} \right)^{-1}\right|,
\end{align}
which is of the order of the light neutrino contributions. Thus, we see that neutrinoless double beta decay does not provide strong bounds in the ISS scenario, since the lepton number violation is suppressed as the scale $\mu$. 

\begin{figure*}[t]

  \includegraphics[width=1.0\textwidth]{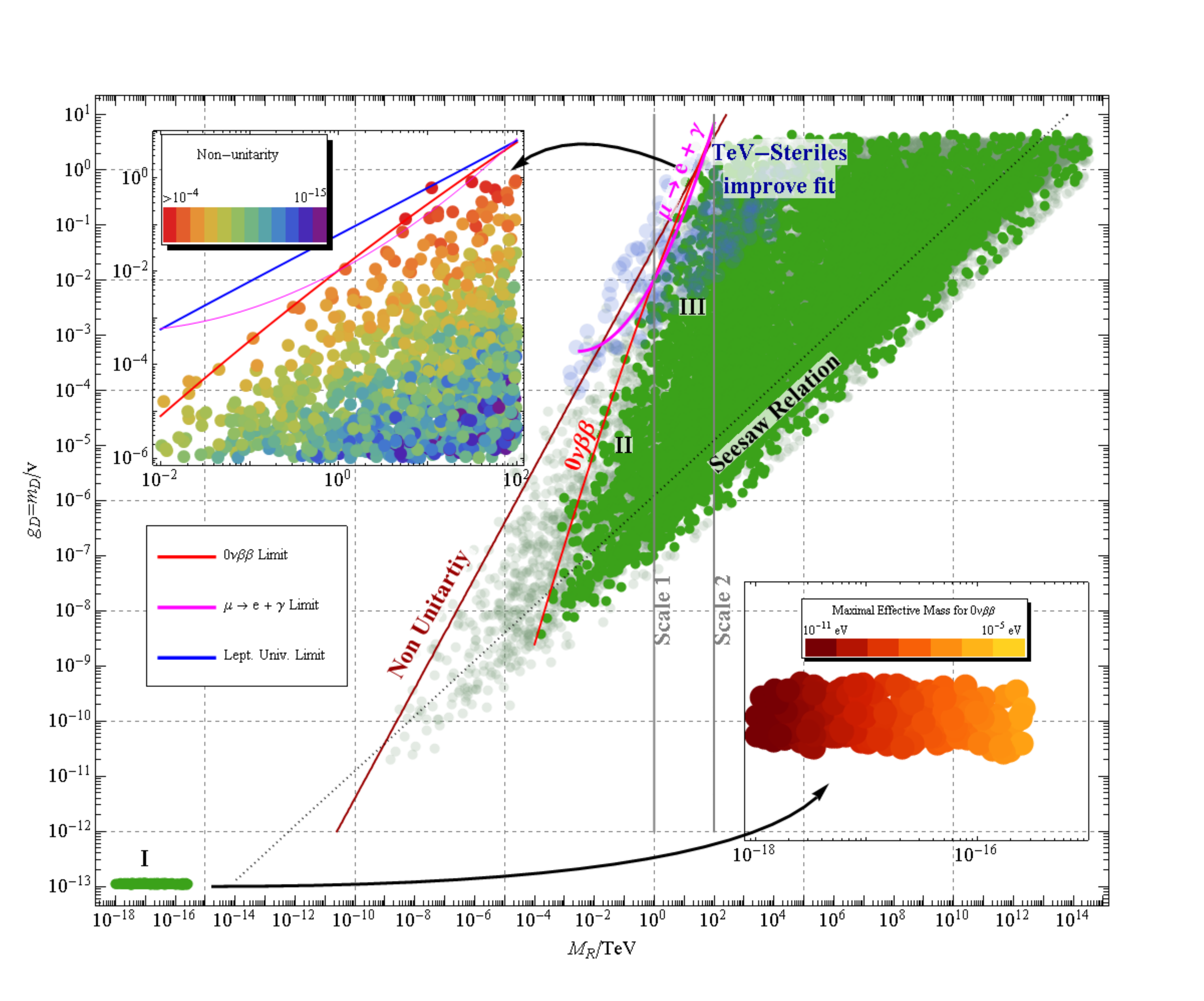}
  \caption{The phenomenologically allowed regions on the Mass Map are displayed. The averaged right handed scale normalized to the TeV scale of symmetry breaking is shown on the x-axis and the averaged EW Dirac mass is shown on the y-axis. Two regions are zoomed in and shown as insets. In the left upper corner the blow up shows the region around the TeV right handed scale, the color coding represents the non-unitartity of the active mixing matrix. In the right lower corner the region of sub-dominant Majorana masses is shown, the color code indicates the maximal expected effective electron neutrino mass for $0\nu\beta\beta$ decays. The experimental constraints are the limits on the rare decay $\mu \rightarrow e + \gamma$, shown as the magenta line, the lepton universality, shown as the blue line and the $0\nu\beta\beta$ decay, leading to the red exclusion line.  The points allowed only in the inverse see-saw are shown in grey-green. The most universal bounds come from non-unitarity constraints, shown as a brown line with gradient one in the log-plot.  The see-saw relation explains the lower boundary of the allowed region, it has gradient one-half in the log-plot. 
\textbf{I} is a fraction of parameter space without a see-saw relation, it is the Pseudo-Dirac region. Here neutrinos come in pairs of strongly mixed left and right particles, with mass splitting induced by Majorana mass fraction.   \textbf{II}  Yukawa see-saw with the upper bound- Scale 1- set by the requirement of perturbative couplings.  \textbf{III} ISS allows perturbative couplings and at the same time the right handed mass up to Scale 2. The most natural parameter choice in the ISS scenario leads to considerable active-sterile mixing of states at the TeV scale and a significantly improved $\chi^2$ of the Electro-Weak fit w.r.t the standard model (light blue points). \hfill} 
 \label{fig:Masterplot}
\end{figure*}

This is not the case for the Lepton flavour violating processes. The best constrained value is the branching ratio  $\text{Br}(\mu \rightarrow e + \gamma )$, where the limit is placed by the MEG collaboration \cite{Adam:2013mnn} and is $5,7 \cdot 10^{-13}$. The neutral fermion contribution to this loop induced decay is
    
\begin{align}
\text{Br}(\mu \rightarrow e + \gamma ) =  \frac{3 \alpha_{\text{em}}}{32 \pi} \left|  \textstyle{\sum_{i}} \mathbf{U}_{\mu i}^* \mathbf{U}_{e i}
\,G \left(\frac{m_i^2}{M_W^2}\right)\right|^2,
\end{align}
where in the loop function $G(x)$ the masses appear squared and the cancellation leading to a vanishing $0\nu\beta\beta$ process can not work.
We find that the MEG bound together with the non-unitarity constraints \cite{Antusch:2006vwa} lead to the most severe constraints on this models, as shown in \fig \ref{fig:Masterplot}. 

As stated before the ISS opens the possibility in the RSSB framework to have states above the TeV scale.  The region of right handed masses between one and a few ten TeV is divided in two subregions, which are distinguished by the value of the active-sterile mixing. 
If this value is sizeable, in fact above $10^{-6}$, the phenomenology is considerably affected. The most sensitive observables are the Z boson invisible decay width and the Muon decay constant, which is used to determine the Fermi constant. The observables dependence on the non unitarity parameters, see \eqn \ref{epsilon} is given by 

\begin{align}
\frac{\Gamma^\text{inv}_Z}{[\Gamma^{\text{inv}}_Z]_\text{SM}} = \frac{1}{3}\sum_\alpha (1-\epsilon_\alpha)^2 \,,\\
G_\mu = G_F (1 - \epsilon_e)(1 - \epsilon_\mu)\,.
\end{align}

This region of seizable active sterile mixing with heavy particles is of particular interest, since here the oblique corrections can become large. So, on the one hand the $\chi^2$ with the EWPOs provides us with phenomenological bounds in this region. On the other hand this is an example of a theory where contributions from heavy sterile neutrinos can improve the electroweak fit, as discussed in \citep{Akhmedov:2013hec}. In \fig \ref{fig:Masterplot} the region with an improved $\chi^2$ is bound towards lower mass values by experimental constraints from the $\mu \rightarrow e + \gamma$ decay and towards higher masses the radiative corrections become incompatible with observations in case of large active-sterile mixing.

Having discussed constraints on the right handed mass, it is interesting to study which Dirac mass scales are allowed. The mass scale of the light neutrinos is set by the following scale relation $m_\nu \approx \mu m_D^2/M_{Rx}^2$, furthermore it is required that $\mu/M_{Rx}=:\delta \ll 1$ and $m_D< M_{Rx}$. Those relations imply that $m_D > \sqrt{m_\nu M_{Rx}/\delta}$. Given a right handed mass scale, the minimal Dirac mass is larger than in the usual see-saw scenario, which implies that the active-sterile mixing has to be larger as well. 

\begin{figure}[h]

  \includegraphics[width=0.45\textwidth]{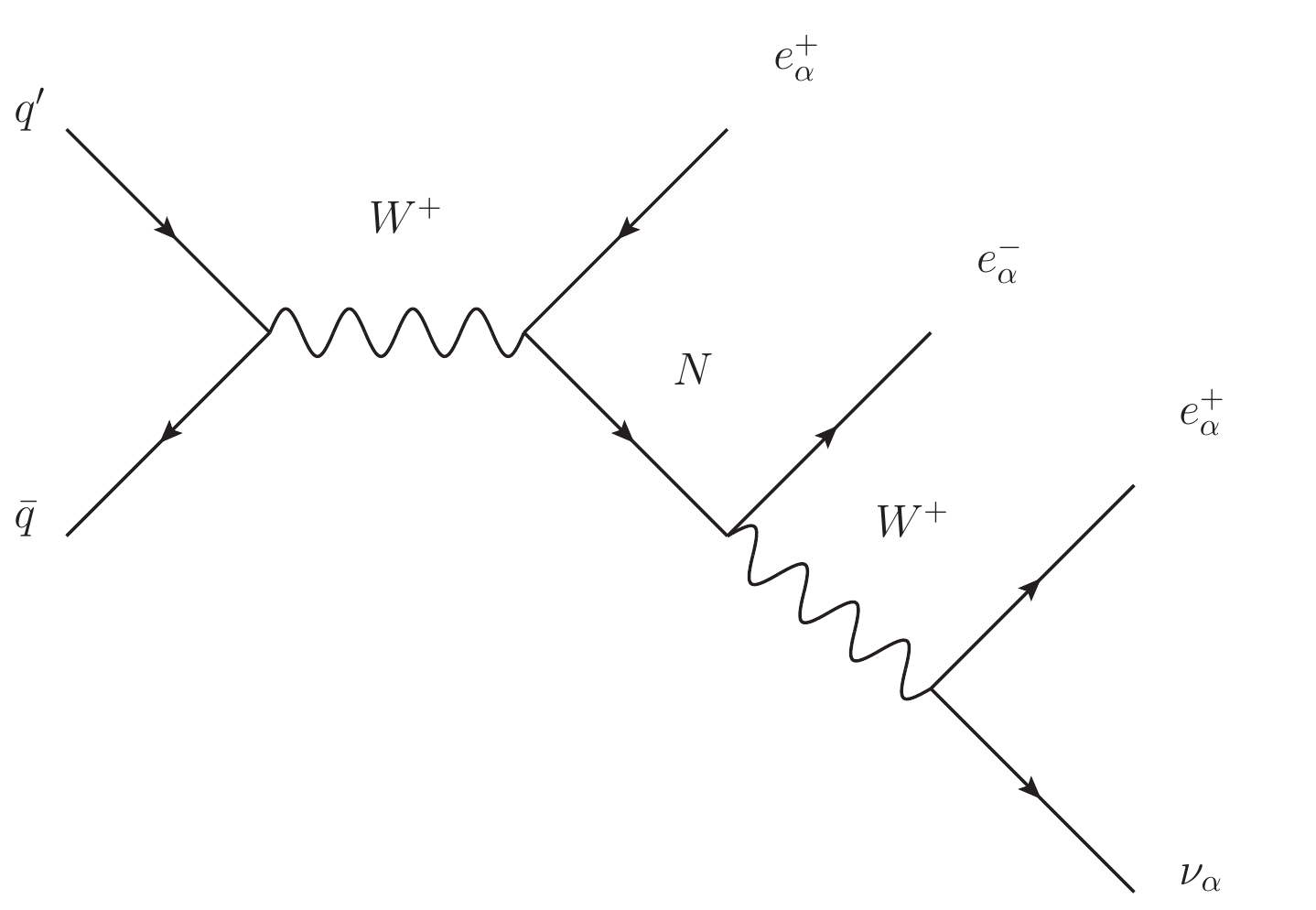}
  \caption{The dominant collider signature for the ISS scenario with the trilepton plus missing energy signature.} 
  \label{fig:DecayDirac}
\end{figure}

The most promising signature to distinguish the heavy Pseudo Dirac neutrino from the ISS scenario from a heavy Majorana neutrino is a direct test at a collider, which is feasible as all the particles involved are around the TeV scale. The difference lies in the dominant decay channel of the right handed neutrinos. While in the Majorana see-saw the lepton number violation is unsuppressed generically, the dominant process is expected to be the lepton number violating decay in \fig \ref{fig:DecayLNV}. In the case of a decay of a heavy Pseudo Dirac neutrino, lepton number violation is suppressed by the smallness of the right handed Majorana scale $\mu$ \cite{Kersten:2007vk}, thus the dominant processes are lepton number conserving decays. As argued in  \cite{Das:2012ze, Das:2014jxa}, the opposite sign dilepton decay has a very large SM background and thus the relevant channel becomes the trilepton decay with missing energy, see \fig \ref{fig:DecayDirac}. As shown by Das et al. the inclusive cross section of the trilepton final state is controlled by the branching ratio of the heavy neutrino in the W boson and a lepton, it has the partial decay width
 \begin{align}
 \label{eq:decay}
\Gamma(N \rightarrow \ell_\alpha W) = \frac{g^2 \,\epsilon_\alpha}{64 \pi} \frac{m_i^3}{M_W^2} \left( 1 - \frac{M_W^2}{m_i^2}\right)^2 \left(1 + 2\frac{M_W^2}{m_i^2} \right).
\end{align}


As shown by \eqn \ref{eq:decay}, the decay width crucially depends on the non-unitarity parameter $\epsilon_\alpha$. The interesting feature of the ISS in the RSSB framework is, that the requirement of no large scale separation results in naturally large active-sterile mixing, as $\epsilon \approx m_D^2/M_{Rx}^2$. Thus the most natural value for $\epsilon$, given an order of magnitude between the scales and Yukawa couplings of order one is about one percent, close to the sensitivity threshold of modern experiments. Note that the recently proposed production mechanism for heavy sterile neutrinos via t-channel processes can further increase the collider sensitivity, as argued in \cite{Dev:2013wba}.

\subsection{Decoupled Hidden Sector}

The discussion of RSSB led us to the finding that generically all scalar scales are close to the TeV scale if no finetuning in the potential is involved. The most natural mechanism to generate the neutrino mass scale, far below was the connection to lepton number violation and thus models seem favourable where this scale is suppressed. We found the most natural model to be the ISS, in this scenario the Majorana scale is generically at the order of keV. We would like to point out that the connection to the dark matter sector in this context seems very promising by considering two set-ups. 
  
Suppose a scenario in which the Hidden sector contains a SM singlet fermion $\nu_x$ with the dark $U(1)$ charge 1 and a SM scalar singlet $\phi_D$ with a dark charge 2 which gets a VEV and thus via the term $\phi_D \bar{\nu}_x \nu_x^c$ generates a mass between the EW and the TeV scale for the fermion. This particle is stable but also almost decoupled from the SM, since the Higgs portal coupling to $\phi_D$ is so far the only allowed interaction channel and it is constrained to be small by experiment. Therefore, it is a decoupled sterile neutrino. It is however possible to switch on a fermionic portal of the form $\bar{\nu}_x \eta \,\ell_R  $, as discussed in \cite{Cao:2009yy, Giacchino:2013bta, Toma:2013bka,  Kopp:2014tsa} and in a scale invariant context in \cite{Antipin:2013exa}. Here $\ell_R$ is a right handed lepton, which is a phenomenologically allowed interaction. The $\eta$ is an electrically charged scalar mediator which has to be of a similar mass due to the requirement of no scalar mass hierarchies. This interaction can lead with the appropriate parameter choice to the production of $\nu_x$ in the early universe with the correct abundance to be a cold dark matter candidate via the lepton portal interaction, as discussed in the literature. This class of models has a rich phenomenology including gamma ray signals which can be peaked and serve as a good DM detection signature \cite{Kopp:2014tsa}. The detailed discussion, however, goes beyond the scope of this work. The intriguing insight is, that the requirement of no scalar mass hierarchy leads automatically to the region of typical WIMP masses.  

In the second scenario the ISS, as in \textbf{1D} and \textbf{2D} with an additional fermionic state $\nu_x$ in the hidden sector, with the charges $(1,0,1)$ in $(SU(2),U(1)_Y, U(1)_\text{Hidden})$ is considered. It is thus a 3 active and 3+3+1 sterile scenario. The mass matrix after eliminating unphysical phases has the structure

\begin{equation}
 \label{eq:seesawInvDM}
        \mathcal{M} = 
        \begin{pmatrix}
          0 & m_D^{3 \times 3} & 0 & 0 \\
          m_D^T & 0 & M_{Rx}^{3 \times 3} & A^{3 \times 1} \\
          0 & M_{Rx}^T &  \mu_1^{3 \times 3} & 0 \\
          0 & A^T &  0 & \mu_2 \\
        \end{pmatrix} \, .
      \end{equation}

The spectrum would be given by  three Pseudo Dirac neutrinos of the scale $M_{Rx}$. The light neutrino mass is given by \eqn \ref{ISSrelations} and with $A \approx M_{Rx}$, which is natural given order one Yukawas, the additional sterile state has a mass of $\mu$ and a small mixing with the active neutrinos of the order $\mu^2/M_{Rx}^2$.  The remarkable feature is that the scale $\mu \approx \text{keV}$ required by the see-saw relation is also the correct scale for this state to be a Dark Matter candidate \cite{Dolgov:2000ew, Bezrukov:2009th}.

We find that incorporating the neutrino mass generation in the RSSB framework naturally provides us with two scales of DM candidates, those are the TeV scale suitable for a cold Dark Matter particle and the keV scale leading to warm Dark Matter.

\section{\label{sec:Conclusion}Conclusion}

We studied in this paper consequences of conformal electro-weak symmetry breaking models for 
neutrino masses. Many phenomenologically viable models contain extra scalars which undergo 
dimensional transmutation. The VEV of this scalar triggers then via the Higgs portal electro-weak 
symmetry breaking. This over-all picture has interesting consequences for neutrino masses.
First, no explicit Dirac or Majorana mass terms are allowed, since they would violate conformal 
symmetry explicitly. All fermion masses must therefore arise as some Yukawa coupling times the
VEV of some scalar. The second generic feature is that Coleman-Weinberg type symmetry 
breaking leads to a loop-generated symmetry breaking effective potential where the Higgs mass
(curvature of the minimum) is loop suppressed compared to the VEV. This can be seen in the 
Standard Model, where Coleman-Weinberg symmetry breaking leads to a Higgs mass of about $9$~GeV,
which is excluded. This explains also why the scale which is generated by the extra scalar should be
in the TeV range in order to match the EW scale. This implies finally that all neutrino masses come
from Dirac and Majorana Yukawa couplings which are multiplied either with the EW scale or with
the TeV-ish symmetry breaking scale of the extra scalar. 

We studied such scenarios in this paper in a rather general context. For that we distinguished 
three basic strategies of accommodating neutrino masses. Embedding of the SM in a larger gauge 
group, enlarging the field content of the SM by additional fields or extending the SM by a Hidden sector. 
In \sect \ref{sec:Models} we present a catalogue of viable conformal neutrino 
mass models and describe them in more detail in the appendix. 

Note that any neutrino mass between zero and the VEVs (or even somewhat bigger) can be obtained 
by selecting the corresponding Yukawa couplings. The wide spectrum of Yukawa couplings for other
fermions of the SM implies that a wide spectrum of neutrino mass terms is expected in these scenarios. 
Note that very tiny neutrino masses are still quite natural, since the discussed models suppress them
via a see-saw or via loops. 
We show that in the  Yukawa see-saw model the adjustment of the couplings can be reduced to the 
same amount as present in the charged lepton sector.  The amount of tuning the Yukawa couplings 
can be largely reduced if the neutrino mass generation is related to lepton number violation. If lepton 
number is taken to be an approximate symmetry, which is broken explicitly in the Lagrangian, the 
smallness of the braking parameter is natural in the sense, that its absence would increase the symmetry. 
The lepton number violation parameters can lead to light neutrino masses via loops, or to small Majorana 
mass contributions of Dirac particle pairs. We present several models where small and or loop suppressed 
lepton number violation is the driving principle behind neutrino mass generation. In particular when 
combined with the ISS mechanism the small Majorana mass fraction in the heavy Dirac neutrino pair 
leads to small active neutrino mass and no fine-tuning is needed.    

In addition we perform  in \sect \ref{sec:Phenomenology} a phenomenological analysis with the goal 
to check whether the models can indeed reproduce the neutrino oscillation data and at the same time 
be consistent with rare decay experiments, as the $0\nu\beta\beta$ searches. Our finding is, that there 
are four phenomenologically viable regions. 

Scenario A has only light neutrinos with Majorana masses, which are generated on tree or loop level by 
particles with lepton number violating couplings. This scenario can lead to detectable signals in the 
$0\nu\beta\beta$ decays and the additional states, such as the triplet scalar can be produced at 
colliders, since their mass must be about the TeV scale.  

Scenario B is a Pseudo-Dirac scenario where pairs of light mass eigenstates are almost degenerate 
with only small Majorana mass fractions. This scenario requires, however, very small Dirac Yukawa 
couplings and is in general experimentally very challenging. The most promising searches for light 
Pseudo Dirac neutrinos are oscillations on cosmic scales which could probe the small mass splitting. 

Scenario C is the Sub-TeV scenario with right handed Majorana states below the TeV scale. This region 
is severely constrained by limits in the $0\nu\beta\beta$ decay, since the lepton number violation is 
unsuppressed. The collider signature which one would expect are decays to same sign dileptons, a 
process practically without SM background.

In scenario D the right handed mass can be up to few ten TeV. This can be achieved in ISS models 
where several scalars are in the game and have a hierarchical VEV structure. The  ISS scenario is of 
particular interest, since it improves the Electro-Weak fit with respect to its SM value. In this parameter 
region the active-sterile mixing is enlarged and can provide testable signals. This conformal ISS is also 
theoretically attractive since it contains Yukawas of order one and the smallness of the hidden sector 
parameters is implied by loop suppression and thus completely avoids fine-tuning. In this region the 
heavy sterile neutrinos are almost mass degenerate Pseudo Dirac pairs with small Majorana mass 
fractions. This leads to a suppression of lepton flavour violation and the most relevant constraints in 
this case, come from searches of lepton flavour violating decays, as $\mu \rightarrow e + \gamma$. 
At colliders a decay of such a heavy neutrino would have a trilepton final state and missing energy 
without lepton number violation as the smoking gun signal. 

We briefly comment that the Hidden sector can contain almost decoupled Dark Matter candidates, which 
can be either coupled via the lepton portal to the SM or due to small active-sterile neutrino mixing. 
The  masses are either at the EW or the keV scale.  The observation here is, that taking the gauge 
hierarchy problem seriously can provide us with a hint for a Dark matter scale.

Additional signals in collider experiments are expected to appear in all viable neutrino mass models, 
since all require new scalar or fermionic states around the TeV mass region. Therefore, we expect that 
all those models can be tested by the LHC. The collider signatures and neutrino experiments combined 
will provide very powerful tools for studying and distinguishing among the different scenarios. 
Phenomenological details of such models and further theoretical aspects will be discussed in future work.

\section*{Acknowledgments}

We would like to thank Hiren Patel, Martin Holthausen, Branimir Radovcic, Julian Heeck and 
Kher-Sham Lim for helpful discussions. 
JS acknowledges support from the IMPRS for Precision Tests of Fundamental Symmetries. 

\appendix
\section{\label{app:Models}Conformal Neutrino Mass Models}
   \subsection*{Models within the SM Gauge Group}
   		We begin with the systematic description of viable conformal neutrino mass models. 
        It will be very important in this section to point out which particles have been integrated out and which
        picture of neutrino mass generation we are considering. 
	\subsubsection*{Models with Dominant Contributions to the Left-Handed Majorana Entry}        
        \begin{itemize}
          \item 
            \underline{  \textbf{3A: SM + $\boldsymbol{\nu^{}_{\!R}}+\boldsymbol{\varphi}$}} 
            \vspace{1mm} \\
            \underline{Particle content}: $L:(2,-1);\; H:(2,1);\; \nu^{}_{\!R}:(1,0);\; 
            \varphi:(1,0)$, \\
            \underline{Yukawa Lagrangian}:
            $-\mathscr{L}_Y=g^{}_{H} \overline{L}\tilde{H}
            \nu^{}_{\!R}+g^{}_{\varphi}\varphi\overline{\nu^{c}_{\!R}}
            \nu^{}_{\!R}$ + h.c. 
            \vspace{1mm} \\
            \underline{Potential}: $V_{\textrm{I}}=\lambda_H(H^\dagger H)^2
            +\lambda_\varphi(\varphi^\dagger\varphi)^2
            +\lambda_{H\varphi}(\varphi^\dagger\varphi)(H^\dagger H)$ 
            \vspace{1mm} \\
            With this we find the diagrams
            \vspace{1cm} \\
            \begin{widetext}
            \begin{fmffile}{conformal1} 
              \begin{eqnarray*}
                \hspace{-2cm} \phantom{+} \hspace{1cm}
                \parbox{45mm}{
                \begin{fmfgraph*}(45,40)
                  \fmftop{X1,S1,S2,S3,X2} \fmflabel{$\langle H\rangle$}{S1} 
                                          \fmflabel{$\langle \varphi\rangle$}{S2}
                  \fmflabel{$\langle H\rangle$}{S3}
        		      \fmfbottom{XP1,P1,P2,P3,XP2}
                  \fmfleft{L1} \fmflabel{L}{L1}
	              \fmfright{L2} \fmflabel{L}{L2}
                  \fmf{fermion,tension=1}{L1,I1}
	              \fmf{fermion,tension=1,label=$\nu^{}_{\!R}$}{I1,I2}
                  \fmf{fermion,tension=1,label=$\nu^{}_{\!R}$}{I3,I2}
                  \fmf{fermion,tension=1}{L2,I3}
                  \fmf{scalar,tension=1}{S1,I1}
                  \fmf{scalar,tension=1}{S2,I2}
                  \fmf{scalar,tension=1}{S3,I3}
                  \fmf{phantom,tension=1}{P1,I1}
                  \fmf{phantom,tension=1}{P2,I2}
                  \fmf{phantom,tension=1}{P3,I3}
                \end{fmfgraph*}
                }
                \hspace{15mm} + \hspace{15mm} 
                \parbox{45mm}{
                \begin{fmfgraph*}(45,40)
                  \fmftop{X1,S1,S2,X2} \fmflabel{$\langle H\rangle$}{S1} 
                                       \fmflabel{$\langle H\rangle$}{S2}
                  \fmfbottom{P1,P2}
                  \fmfbottom{S3} \fmflabel{$\langle\varphi\rangle$}{S3}
      			  \fmftop{P3}
      			  \fmfleft{L1} \fmflabel{L}{L1}
      			  \fmfright{L2} \fmflabel{L}{L2}
      			  \fmf{fermion,tension=2}{L1,I1}
     		      \fmf{fermion,tension=1,label=$\nu^{}_{\!R}$}{I1,I2}
    	 			  \fmf{fermion,tension=1,label=$\nu^{}_{\!R}$}{I3,I2}
     			  \fmf{fermion,tension=2}{L2,I3}
   			      \fmf{scalar,tension=1}{S1,V}
  			      \fmf{scalar,tension=1}{S2,V}
       			  \fmf{scalar,right=0.5,label=H}{V,I1}
      			  \fmf{scalar,left=0.5,label=H}{V,I3}
		          \fmf{scalar}{S3,I2}
  			      \fmf{phantom,tension=1}{VP,P1}
  			      \fmf{phantom,tension=1}{VP,P2}
 			      \fmf{phantom,left=0.5}{I1,VP}
 			      \fmf{phantom,right=0.5}{I3,VP}
 			      \fmf{phantom}{I2,P3}
  		        \end{fmfgraph*}
                }
                \end{eqnarray*}
      		    \vspace{8mm}
                \begin{eqnarray*}
     		      \hspace{-2cm} + \hspace{1cm}
      			  \parbox{45mm}{
    			    \begin{fmfgraph*}(45,40)
    			    \fmftop{X4,S1,S2,S3,X3} \fmflabel{$\langle H\rangle$}{S1} 
                                        \fmflabel{$\langle\varphi\rangle$}{S2}
                        				   \fmflabel{$\langle H\rangle$}{S3}
                \fmfbottom{XP4,P1,P2,P3,XP3}
                \fmfleft{L1} \fmflabel{L}{L1}
                \fmfright{L2} \fmflabel{L}{L2}
                \fmf{fermion,tension=3}{L1,I1}
                \fmf{fermion,tension=1,label=$\nu^{}_{\!R}$}{I1,I2}
                \fmf{fermion,tension=2.5,label=$\nu^{}_{\!R}$}{I3,I2}
                \fmf{fermion,tension=2.5}{L2,I3}
                \fmf{scalar,tension=1}{S1,V}
                \fmf{scalar,tension=1}{S2,V}
                \fmf{scalar,right=0.5,tension=1,label=H}{V,I1}
                \fmf{scalar,left=0.5,tension=1,label=$\varphi$}{V,I2}
                \fmf{scalar}{S3,I3}
                \fmf{phantom,tension=1}{VP,P1}
                \fmf{phantom,tension=1}{VP,P2}
                \fmf{phantom,left=0.5,tension=1}{I1,VP}
                \fmf{phantom,right=0.5,tension=1}{I2,VP}
                \fmf{phantom}{I3,P3}
              \end{fmfgraph*}
              }
              \hspace{15mm} + \hspace{15mm} 
              \parbox{45mm}{
              \begin{fmfgraph*}(45,40)
                \fmftop{X1,S1,S2,S3,X2} \fmflabel{$\langle H\rangle$}{S1}  
                                        \fmflabel{$\langle\varphi\rangle$}{S2}
                                        \fmflabel{$\langle H\rangle$}{S3}
                \fmfbottom{XP1,P1,P2,P3,XP2}
                \fmfleft{L1} \fmflabel{L}{L1}
                \fmfright{L2} \fmflabel{L}{L2}
                \fmf{fermion,tension=2.5}{L1,I1}
                \fmf{fermion,tension=2.5,label=$\nu^{}_{\!R}$}{I1,I2}
                \fmf{fermion,tension=1,label=$\nu^{}_{\!R}$}{I3,I2}
                \fmf{fermion,tension=3}{L2,I3}
                \fmf{scalar,tension=1}{S2,V}
                \fmf{scalar,tension=1}{S3,V}
                \fmf{scalar,right=0.5,tension=1,label=$\varphi$}{V,I2}
                \fmf{scalar,left=0.5,tension=1,label=H}{V,I3}
                \fmf{scalar}{S1,I1}
                \fmf{phantom,tension=1}{VP,P2}
                \fmf{phantom,tension=1}{VP,P3}
                \fmf{phantom,left=0.5,tension=1}{I2,VP}
                \fmf{phantom,right=0.5,tension=1}{I3,VP}
                \fmf{phantom}{I1,P1}
              \end{fmfgraph*}
              }
              \end{eqnarray*}        
            \end{fmffile}
			
			\end{widetext}            
            
            The first diagram is the tree level contribution while the other three are one-loop
            corrections to the first diagram and have thus a smaller contribution to the total
            neutrino mass. Further contributions have either at least two loops or 9 
            mass insertions and thus have even smaller impact on the masses. 
            
            The mass matrix has the following structure:

  			\begin{equation}
              \mathcal{M} = 
              \begin{pmatrix}
                M_L & m_D \\
                m_D & M_R
              \end{pmatrix} \, .
            \end{equation}
            
            The masses are given by $M_R = g_\phi \ev{\varphi}$,
            $m_D = g_h \ev{H}$ and the loop supressed left handed contributions 
            \begin{align}
            M_L \approx \frac{g_H^2 g_\phi \ev{H}^2\ev{\varphi}}{ (4\pi)^2 \Lambda^2}+ \frac{g_H^2 g_\phi \ev{H}^2\ev{\varphi}}{M_R (4\pi)^2 \Lambda}
            \end{align}
            with $\Lambda$ the 
            dominant loop mass contribution. Integrating out the heavier right handed states
            leads to an effective mass for the light species of the order 
            
            \begin{align}
            m_\nu \approx g_H^2 \frac{\ev{H}^2}{M_R} +  \frac{g_H^2 g_\phi \ev{H}^2\ev{\varphi}}{ (4\pi)^2 \Lambda^2}+ \frac{g_H^2 g_\phi \ev{H}^2\ev{\varphi}}{M_R (4\pi)^2 \Lambda} \\ \nonumber = \frac{g_H^2}{g_\phi} \frac{v_H^2}{v_\varphi} \left( 1 + \frac{g_\phi v_\varphi}{(4\pi)^2\Lambda} +   \frac{g_\phi^2 \,v_\varphi^2 }{ (4\pi)^2 \Lambda^2}\right).
            \end{align}
            With $g_H$ being the Dirac and $g_\phi$ the Majorana type Yukawa coupling.
            The tree level contribution dominates in this scenario. We refer to this model type 
            as the Yukawa see-saw.
            
            We consider now in addition a moled with tree level correction to the left-handed 
            Majorana entry by introducing a scalar triplet.

          \item
            \underline{\textbf{5A: SM + $\boldsymbol{\Delta}$ + $\boldsymbol{\varphi}$}} 
            \vspace{1mm} \\
            \underline{Particle content}: $L:(2,-1);\; H:(2,1);\; \Delta:(3,-2);\; \varphi:(1,0)$
            \vspace{1mm} \\
            \underline{Yukawa Lagrangian}: $-\mathcal{L}_Y=g_\Delta\bar{L}\vec{\sigma}\Delta L^c+h.c.
            =g_\Delta(\bar{L}\vec{\sigma}\Delta L^c+\bar{L}^c\vec{\sigma}\Delta^*L)$
            \vspace{1mm} \\
            \underline{Potential}:              
            \begin{align}
              \nonumber V_{\textrm{II}}= \lambda_H (H^\dagger H)^2 + \lambda_{\Delta T}Tr(\Delta^\dagger \Delta)^2     + \lambda_{T \Delta}(Tr(\Delta^\dagger \Delta))^2  \\ \nonumber
          + \lambda_{H\Delta, 1}(H^\dagger H) Tr\Delta^\dagger \Delta + \lambda_{H\Delta,2} H^\dagger\Delta \Delta^\dagger H  \\ \nonumber
              +\lambda_\varphi(\varphi^\dagger \varphi)^2
              +\lambda_{H\varphi}(\varphi^\dagger \varphi)(H^\dagger H) \\ \nonumber
              + \lambda_{\varphi\boldsymbol{\Delta}}(\varphi^\dagger
              \varphi)\Tr{\boldsymbol{\Delta}^\dagger
              \boldsymbol{\Delta}}+\lambda_{\varphi\boldsymbol{\Delta} H}[\varphi H^Ti\sigma_2  
              \boldsymbol{\Delta} H+h.c.] .
            \end{align}   
        All 1-Particle-Irreducible (1PI) diagrams with upto 3 mass insertions and maximum one loop are given by
            \vspace{8mm} \\
         
 			\begin{widetext}
            \begin{fmffile}{conformalJuri2} 
              \begin{eqnarray*}
                \hspace{1cm} \phantom{+} 
                 \parbox{45mm}{
				\begin{fmfgraph*}(45,40)
                  \fmfleft{L1} \fmflabel{L}{L1}
                  \fmfright{L2} \fmflabel{L}{L2}
                  \fmftop{S1} \fmflabel{$\langle \Delta \rangle$}{S1}
                  \fmfbottom{P1}
                  \fmf{fermion}{L1,I1}
                  \fmf{fermion}{L2,I1}
                  \fmf{scalar}{I1,S1}
                  \fmf{phantom}{I1,P1}
                \end{fmfgraph*}                
                  }
                \hspace{5mm} +  \hspace{5mm} 
                \parbox{45mm}
                {
                \begin{fmfgraph*}(45,40)
                  \fmfleft{L1} \fmflabel{L}{L1}
                  \fmfright{L2} \fmflabel{L}{L2}
                  \fmftop{X1,S1,S2,X2} \fmflabel{$\langle\Delta\rangle$}{S1} 
                                       \fmflabel{$\langle\Delta\rangle$}{S2}                          
                  \fmfbottom{S3} \fmflabel{$\langle\Delta\rangle$}{S3}    
                  \fmfbottom{XP1,P1,P2,XP2}
                  \fmftop{P3} 
                  \fmf{fermion,tension=2.5}{L1,I1}
                  \fmf{fermion,label=L,tension=1}{I2,I1}
                  \fmf{fermion,label=L,tension=1}{I2,I3}
                  \fmf{fermion,tension=2.5}{L2,I3}
                  \fmf{scalar,left=0.5,label=$\Delta$}{I1,V}
                  \fmf{phantom,left=0.5}{I1,VP}
                  \fmf{scalar,right=0.5,label=$\Delta$}{I3,V}
                  \fmf{phantom,right=0.5}{I3,VP}
                  \fmf{scalar}{V,S1}
                  \fmf{phantom}{VP,P1}
                  \fmf{scalar}{V,S2}
                  \fmf{phantom}{VP,P2}
                  \fmf{scalar}{S3,I2}
                  \fmf{phantom}{P3,I2}
                \end{fmfgraph*}
                }
                  \hspace{5mm} +  \hspace{5mm} 
                 \parbox{45mm}
                {
                \begin{fmfgraph*}(40,50)
                  \fmfleft{L1} \fmflabel{L}{L1}
                  \fmfright{L2} \fmflabel{L}{L2}
                  \fmftop{X1,S1,S2,S3,X2} \fmflabel{$\langle H\rangle$}{S1} 
                                          \fmflabel{$\langle\varphi\rangle$}{S2}
                                          \fmflabel{$\langle H\rangle$}{S3}
                  \fmfbottom{PX1,P1,P2,P3,PX2}
                  \fmf{fermion}{L1,I1}
                  \fmf{fermion}{L2,I1}
                  \fmf{scalar,label=$\Delta$,tension=2.5}{I1,V}
                  \fmf{phantom,tension=2.5}{I1,VP}
                  \fmf{scalar}{S1,V}
                  \fmf{phantom}{P1,VP}
                  \fmf{scalar}{S2,V}
                  \fmf{phantom}{P2,VP}
                  \fmf{scalar}{S3,V}
                  \fmf{phantom}{P3,VP}
                \end{fmfgraph*}
                }  
                \end{eqnarray*}
              
            \end{fmffile}
			
			\end{widetext}

            The theory at hand is the conformal analogue of the type II see-saw mechanism. 
            Based on measurments of EWPOs the VEV $\langle \Delta_0 \rangle$ has to be orders 
            of magnitude below the EW scale and in our single scale scenario it seems more natural for
            it to be exactly zero at tree level. Therefore, the main contribution comes from the third 
            diagram which yields the neutrino mass
            \begin{align}
              M_L = g_\Delta \frac{\lambda_{\varphi \Delta H}}{M_\Delta^2} \langle \varphi \rangle
              \langle H \rangle^2,
            \end{align}
            where $M_\Delta$ is the physical mass of the scalar triplet. This is controlled by the lepton number violating coupling $\lambda_{\varphi \Delta H} $, furthermore the neutrino mass is suppressed by the mass of the triplet scalar. The mass of the double charged triplet component is experimentally constrained to be above 450 GeV \cite{CMS:2012ulp} and since there should be no large splitting among the components we assume the neutral component to be at least of the same order.

          This model can be enlarged by right handed neutrinos, which leads us to  
            
          \item
            \underline{\textbf{6A: SM+$\boldsymbol{\nu^{}_{\!R}}$+$\boldsymbol{\varphi}$
            +$\boldsymbol{\Delta}$ }} 
            \vspace{1mm} \\
            \underline{Particle content}: $L:(2,-1);\; H:(2,1);\; \Delta:(3,-2);\; \varphi:(1,0);\;
            \nu^{}_{\!R}:(1,0)$
            \vspace{1mm} \\
            \underline{Yukawa Lagrangian}: $-\mathscr{L}_Y=g^{}_{H}\bar{L}\tilde{H}\nu^{}_{\!R}
            +g^{}_{\varphi}\varphi\bar{\nu}^{c}_{\!R}\nu^{}_{\!R}
            +g_\Delta\bar{L}\vec{\sigma}\Delta L^c+h.c.$
            \vspace{1mm} \\
            \underline{Potential}: $V=V_{\textrm{II}}$ 
            \vspace{1mm} \\
            The following diagram is additional to those of \textbf{3A}
            and \textbf{5A}
            
                \begin{fmffile}{conformalJuri3} 
              \begin{eqnarray*}
                \hspace{1cm} \phantom{+} 
                 \parbox{45mm}{
				\begin{fmfgraph*}(45,40)
                  \fmftop{X4,S1,S2,S3,X3} 
                  \fmflabel{$\langle H\rangle$}{S1} 
                  \fmflabel{$\langle\Delta\rangle$}{S2}
                  \fmflabel{$\langle H\rangle$}{S3}
                  \fmfbottom{XP4,P1,P2,P3,XP3}
                  \fmfleft{L1} \fmflabel{L}{L1}
                  \fmfright{L2} \fmflabel{L}{L2}
                  \fmf{fermion,tension=3}{L1,I1}
                  \fmf{fermion,tension=1,label=$\nu^{}_{\!R}$}{I1,I2}
                  \fmf{fermion,tension=2.5,label=$\nu^{}_{\!R}$}{I3,I2}
                  \fmf{fermion,tension=2.5}{L2,I3}
                  \fmf{scalar,tension=1}{V,S1}
                  \fmf{scalar,tension=1}{V,S2}
                  \fmf{scalar,right=0.5,tension=1,label=H}{V,I1}
                  \fmf{scalar,left=0.5,tension=1,label=$\varphi$}{V,I2}
                  \fmf{scalar}{S3,I3}
                  \fmf{phantom,tension=1}{VP,P1}
                  \fmf{phantom,tension=1}{VP,P2}
                  \fmf{phantom,left=0.5,tension=1}{I1,VP}
                  \fmf{phantom,right=0.5,tension=1}{I2,VP}
                  \fmf{phantom}{I3,P3}
                \end{fmfgraph*}                
                  }
             \end{eqnarray*}
             \end{fmffile}

            The diagram contributes to the left handed mass an approximate term of the order
            $\ev{H}^2\ev{\Delta}/((4\pi)^2 \Lambda \,M_R)$ which is supressed by the smallness of the triplet 
            VEV and therefore subdominant. In this model the $\varphi$ field can have a VEV, which brings us to the Yukawa see-saw scenario, or it can have no VEV and the right handed neutrino only adds a Dirac contribution to the neutrino mass. In this case the phenomenology would be of the Pseudo Dirac scenario.

            Like seen in the non-conformal case it is also possible to introduce a triplet fermion
            to couple to the left-handed doublet. Unlike in the non-conformal scenario we now
            have to introduce an uncharged singlet scalar to generate neutrino masses.

          \item 
            \underline{\textbf{10A: SM + $\boldsymbol{\Sigma}+\boldsymbol{\varphi}$}} 
            \vspace{1mm} \\
            \underline{Particle content}: $L:(2,-1);\; H:(2,1);\; \Sigma:(3,0);\; 
            \varphi:(1,0)$, \\
            \underline{Yukawa Lagrangian}:
            $-\mathscr{L}_Y=g_\Sigma \tilde{H}^\dagger \overline{\Sigma} L + 
            g_\varphi \varphi \Tr{\left[ \overline{\Sigma^c} \Sigma \right]} + h.c. $ 
            \vspace{1mm} \\
            \underline{Potential}: $V = V_{\textrm{I}}$ 
            \vspace{1mm} \\
            The main contribution to the neutrino mass is given by
            \vspace{5mm} \\
            \begin{fmffile}{conformal4}
              \begin{eqnarray*}
                \parbox{55mm}{
                \begin{fmfgraph*}(55,40)
                  \fmfstraight
                  
                  \fmfleft{L} \fmflabel{L}{L}  
                  \fmfright{R} \fmflabel{L}{R}
                  \fmftop{P1,H1,F,H3,P2} \fmflabel{$\langle H \rangle$}{H1}
                                   \fmflabel{$\langle \varphi \rangle$}{F}
                                   \fmflabel{$\langle H \rangle$}{H3}
                                   
                  \fmf{fermion}{L,V1}
                  \fmf{fermion,label=$\Sigma$}{V1,V2}
                  \fmf{fermion,label=$\Sigma$}{V3,V2}
                  \fmf{fermion}{R,V3}
                  
                  \fmffreeze
                  
                  \fmf{scalar}{H1,V1}
                  \fmf{scalar}{F,V2}
                  \fmf{scalar}{H3,V3}
                \end{fmfgraph*}            
                }
              \end{eqnarray*}
            \end{fmffile}
            This diagram yields the mass
            \begin{equation}
              M_L = g_\Sigma^2 \frac{\langle H \rangle^2}{g_\varphi \langle \varphi \rangle}.
            \end{equation}

 \subsubsection*{Models with Dominant Contributions to the Right-Handed Majorana Entry}
      Already in model \textbf{3A} right handed neutrinos with Majorana mass were considered.
      There are, however, further ways to influence the right-handed Majorana mass. The first 
      possibility we want to study is to introduce a scalar and a fermion triplet and a scalar
      singlet. 

      \item 
        \underline{\textbf{1B: SM + $\boldsymbol{\nu_R} + \boldsymbol{\Sigma}+\boldsymbol{\Delta}+\boldsymbol{\varphi}$}} \\
        \underline{Particle content}: $L:(2,-1) ;\; \nu_R:(1,0);\; \Sigma:(3,0);\; H:(2,1);\;  \Delta:(3,0);\;
        \varphi:(1,0)$      
        \vspace{1mm} \\
        \underline{Yukawa Lagrangian}: $-\mathscr{L}_Y =g^{}_{H}\bar{L}\tilde{H}\nu^{}_{\!R} + g_{\Delta} \Tr{[\overline{\Sigma} 
        \boldsymbol{\Delta}
        \nu_R]} + g_{\varphi,1}\Tr{[\varphi \overline{\Sigma^ c}\Sigma]} + 
        g_{\varphi,2} \varphi \overline{\nu_R} \nu_R^c + h.c.$
        \vspace{1mm} \\
        The relevant lepton number violating term in the potential is displayed.
         
        \underline{Potential}: $V \supset \lambda \varphi H^Ti\sigma_2\boldsymbol{\Delta}^\dagger \tilde{H}
        + h.c.$
        \vspace{1mm} \\
        Furthermore we forbid the VEV of $\Delta$.
        In addition to the diagram of \textbf{3A} we get the diagram
        \vspace{8mm} \\
 		\begin{fmffile}{conformalJuri5}
              \begin{eqnarray*}
                \parbox{55mm}{
                \begin{fmfgraph*}(55,40)
                   \fmfstraight
              \fmfleft{L} \fmflabel{$\nu_R$}{L}
              \fmfright{R} \fmflabel{$\nu_R$}{R}
              \fmftop{A,T1,T2,T3,T4,T5,T6,T7,B} \fmflabel{$\langle H \rangle$}{T1}
                                            \fmflabel{$\langle \varphi \rangle$}{T2}
                                            \fmflabel{$\langle H \rangle$}{T3}
                                            \fmflabel{$\langle H \rangle$}{T5}
                                            \fmflabel{$\langle \varphi \rangle$}{T6}
                                            \fmflabel{$\langle H \rangle$}{T7}
              \fmfbottom{xA,xT1,xT2,xT3,xT4,xT5,xT6,xT7,xB}
              \fmflabel{$\langle \varphi \rangle$}{xT4}
                                            
              \fmf{fermion}{L,V1}
              \fmf{fermion,label=$\Sigma$}{V1,V2}
              \fmf{fermion,label=$\Sigma$}{V3,V2} 
              \fmf{fermion}{R,V3}
              
              \fmffreeze
              
              \fmf{scalar,label=$\Delta$,tension=2.5}{TV1,V1}
              \fmf{scalar,tension=2.5}{xT4,V2}
              \fmf{scalar,label=$\Delta$,tension=2.5}{TV3,V3}
              
              \fmf{scalar}{T1,TV1}
              \fmf{scalar}{T2,TV1}
              \fmf{scalar}{TV1,T3}
              
              \fmf{scalar}{T5,TV3}
              \fmf{scalar}{T6,TV3}
              \fmf{scalar}{TV3,T7}
                \end{fmfgraph*}            
                }
              \end{eqnarray*}
            \end{fmffile}        

        \vspace{1cm} 
        Note that the scalar triplet $\Delta$ cannot be used to generate left-handed Majorana masses
        as it has the wrong hypercharge. Adding contributions from both diagrams the right-handed mass is given by
        \begin{equation}
          \begin{split}
          M_R & = g_{\varphi,2} \langle \varphi \rangle + \lambda^2 g_{\Delta}^2 
          \frac{\langle H \rangle^4 \langle \varphi \rangle^2}{M_\Sigma \cdot M_\Delta^4} \\
          & \approx \left( g_{\varphi,2} +  g_{\Delta}^2 \frac{\text{GeV}^2}{g_{\varphi,1} \ev{\varphi}^2}
          \right) \ev{ \varphi }.
          \end{split} 
        \end{equation}
      
        Here the fact was used, that the combination $\lambda \ev{H}^2\ev{\varphi}/M_\Delta$ from the diagram induces an effective VEV of the triplet field , which is experimentally constrained by measurements of the $\rho$ parameter to be $\ev{\Delta} \lesssim 1 \text{GeV}$. Thus the second term is subdominant.
        
        \vspace*{3mm}
        
\subsection*{Models with an Additional Hidden Sector Symmetry}
 The particle content is extended by additional SM singlet fermions. However, those would not be distinguishable from the sterile neutrinos 
  $\nu_R$ if they had all quantum numbers in common. Now with the Hidden Sector symmetry, which will be
    denoted by $U(1)_H$, there are observable effects. The SM singlet fermions with a hidden charge are denoted by $\nu_x$ and this requires the mass matrix    to be extended to $3 \times 3$ in the one flavour case
    \begin{equation}
      \mathcal{M} = 
      \begin{pmatrix}
        M_L & m_D & 0 \\
        m_D & M_R & M_{Rx} \\
        0 & M_{Rx} & M_x 
      \end{pmatrix} \, .
    \end{equation}
 	 Note that the sterile neutrino $\nu_R$ must not carry a hidden charge, as otherwise coupling to the Higgs would be forbidden and
 	 the complete sector would decouple.

\subsubsection*{Modifying the $\nu_R$ Majorana Mass}
     We begin with a theory in which the direct term
      \begin{equation}
        g \varphi \overline{\nu_R} \nu_R^c
      \end{equation}
      is forbidden by the additional HS symmetry.
    
        \item
         \underline{\textbf{1C: SM  $ \times \,\boldsymbol{U_H(1)} $}} \\
          \underline{Particle content}: $L:(2,-1, 0) ;\; H:(2, 1,0) ;\; \nu_R:(1,0,0);\; \nu_x:(1,0,1);\; \varphi_1:(1,0,1);\;
          \varphi_2:(1,0,2)$,
          \vspace{1mm} \\
          where the third number in brackets denotes the HS charge. This particle content yields
          the additional terms 
          \vspace{1mm} \\
          \underline{Yukawa Lagrangian}: $-\mathscr{L}_{Y_1} = g_1 \varphi_1 \overline{\nu_R} \nu_x^c + 
          g_2 \varphi_2 \overline{\nu_x} \nu_x^c +g^{}_{H}\bar{L}\tilde{H}\nu^{}_{\!R}$
          \vspace{1mm} \\
          If $\varphi_1$ and $\varphi_2$ get a VEV this theory yields the mass matrix
          \begin{equation}
            \mathcal{M} = 
            \begin{pmatrix}
              0 & m_D & 0 \\
              m_D & 0 & M_{Rx} \\
              0 & M_{Rx} & M_x
            \end{pmatrix} \, .
          \end{equation}      
          This mass matrix represents the double see-saw mechanism \cite{Barr:2003nn}.
         In language of diagrams this model is represented by
          \vspace{5mm} \\
          \begin{fmffile}{double_seesaw}
            \begin{eqnarray*}
              \parbox{55mm}{
              \begin{fmfgraph*}(55,45)
              \fmfstraight
                \fmfleft{L} \fmflabel{$\nu_R$}{L}
                \fmfright{R} \fmflabel{$\nu_R$}{R}
                \fmftop{A,T1,T2,T3,B} \fmflabel{$\langle \varphi_1 \rangle$}{T1}
                                  \fmflabel{$\langle \varphi_2 \rangle$}{T2}
                                  \fmflabel{$\langle \varphi_1 \rangle$}{T3}
                                  
                \fmf{fermion}{L,V1}
                \fmf{fermion,label=$\nu_x$}{V2,V1}
                \fmf{fermion,label=$\nu_x$}{V2,V3}
                \fmf{fermion}{R,V3}
                
                \fmffreeze
                
                \fmf{scalar}{V1,T1}
                \fmf{scalar}{T2,V2}
                \fmf{scalar}{V3,T3}
              \end{fmfgraph*}
              }
            \end{eqnarray*}
          \end{fmffile}
          \\
         Integrating out $\nu_x$ we obtain an effective mass $M_R$ and find the contracted mass matrix
          \begin{equation}
            \mathcal{M} = 
            \begin{pmatrix}
              0 & m_D \\
              m_D & M_R
            \end{pmatrix} \, ,
          \end{equation}
          where $M_R$ can be calculated from the diagram. Two cases are relevant, either if $M_{Rx}<<M_x$ one has
          \begin{equation}
            M_R \approx \frac{g_1^2}{g_2}
            \frac{\langle \varphi_1 \rangle^2}{\langle \varphi_2 \rangle} \, ,
          \end{equation}
or in the other limit $M_{Rx}>>M_x$  the mass is 
          \begin{equation}
            M_R \approx M_{Rx} = g_1 \langle \varphi_1 \rangle \, .
          \end{equation}
          
          This indicates that it is possible to have either the double or the inverse see-saw scenario realized. So far there is no reason to assume that $M_x$ is small, thus the more natural scenario in this model is the double see-saw, leading to a Sub-TeV see-saw scenario.

		\item
           \underline{\textbf{2C: SM  $ \times \,\boldsymbol{U_H(1)} $}} \\
           \underline{Particle content}: $ L:(2,-1, 0) ;\; H:(2,1,0) ;\; \nu_R:(1,0,0);\; \nu_x:(1,0,2);\; \varphi_1:(1,0,0) ;\; \varphi_2:(1,0,-2)$          
            \vspace{1mm} \\
          \underline{Yukawa Lagrangian}: $-\mathscr{L}_Y \supset -\mathscr{L}_{Y_1} $
        
         We see that the Majorana mass term for the hidden sector fermion can not be constructed and hence the matrix structure is 
          \begin{equation}
            \mathcal{M} = 
            \begin{pmatrix}
              0 & m_D & 0 \\
              m_D & M_R & M_{Rx} \\
              0 & M_{Rx} &0
            \end{pmatrix} \, .
          \end{equation}  
                
      This is a structure of the minimal extended see-saw, discussed in \cite{Heeck:2012bz}, but here it is at the TeV scale.  The interesting feature is that with $M_R > m_D$ and $M_R > M_{Rx}$ this	 see-saw scenario generates light active and sterile neutrinos which can have large mixing with the active sector. The light sterile neutrino could for instance explain the missing upturn in the Super Kamiokande data, as discussed in \cite{Smirnov:2014zga}, a detailed discussion of this scenario is beyond the scope of this work.
       
       A phenomenologically different scenario occurs if we forbid the VEV $\langle \varphi_1
       \rangle$. Consider the following theory.

         \item
           \underline{\textbf{3C: SM  $ \times \,\boldsymbol{U_H(1)} $}} \\
           \underline{Particle content}: $ L:(2,-1,0) ;\; H:(2,1,0) ;\;\nu_R:(1,0,0);\; \nu_x:(1,0,1);\; \varphi_1:(1,0,1);\;
           \varphi_2:(1,0,2);\; \varphi_3:(1,0,-4)$,
           \vspace{1mm} \\
           Note that the newly introduced SM singlet scalar $\varphi_3$ does not change the Yukawa
           Lagrangian. There is, however, an additional potential term:
           \vspace{1mm} \\
           \underline{Potential}: $V \supset \lambda \varphi_1^2 \varphi_2 \varphi_3 \, + h.c.$
           \vspace{1mm} \\
           Thus if we forbid, as mentioned, the VEV of $\varphi_1$, the diagram with the main
           contribution to $M_R$ is given by
           \vspace{7mm} \\
           \begin{fmffile}{right_radiative}
             \begin{eqnarray*}
               \parbox{55mm}{
               \begin{fmfgraph*}(55,45)
                 \fmfleft{L} \fmflabel{$\nu_R$}{L}
                 \fmfright{R} \fmflabel{$\nu_R$}{R}
                 \fmftop{TA,T1,T2,TB} \fmflabel{$\langle \varphi_2 \rangle$}{T1}
                                    \fmflabel{$\langle \varphi_3 \rangle$}{T2}
                 \fmfbottom{BA,BB,B,BC,BD} \fmflabel{$\langle \varphi_2 \rangle$}{B}
                 
                 \fmf{fermion}{L,V1}
                 \fmf{fermion,label=$\nu_x$}{V2,V1}
                 \fmf{fermion,label=$\nu_x$}{V2,V3}
                 \fmf{fermion}{R,V3}
                 
                 \fmffreeze
                 
                 \fmf{scalar,left=.5,label=$\varphi_1$}{V1,TV}
                 \fmf{scalar,right=.5,label=$\varphi_1$}{V3,TV}
                 \fmf{scalar,tension=1.3}{T1,TV}
                 \fmf{scalar,tension=1.3}{T2,TV}
                 \fmf{scalar}{B,V2}
               \end{fmfgraph*}
               }
             \end{eqnarray*}
           \end{fmffile}
           \vspace{8mm} \\
         The mass of the right handed neutrino is generated at one loop and the effective mass matrix reads
           \begin{equation}
             \mathcal{M} = 
             \begin{pmatrix}
               0 & m_D \\
               m_D & M_R
             \end{pmatrix} \, .
           \end{equation}
           To approximate the scale of $M_R$ we use the fact that this loop has the same topology as in the Ma model. Therefore, the right handed mass scale is 
   
   			\begin{align}
             M_R \approx  \frac{\lambda}{16\pi^2} \frac{ g_1^2}{g_2} \ev{\varphi_3} I\left(\frac{\ev{\varphi_2}^2}{M_{\varphi_1}^2} \right)  \, , \\ 
             \text{ with   } I(x) = \frac{1}{1-x}\left(1+\frac{x \log{x}}{1-x}\right)  \,.            
           \end{align}      
   			Thus the right handed mass is loop suppressed and controlled by the parameter $\lambda$, which if set to zero increases the Lagrangian symmetry. Therefore, this model leads to a scenario with Pseudo-Dirac active neutrinos.

         \item
           \underline{\textbf{4C: SM  $ \times \,\boldsymbol{U_H(1)} $}} \\
           \underline{Particle content}: $L:(2,-1,0) ;\; H:(2,1,0) ;\; \nu_R:(1,0,0); \; \nu_x:(1,0,1); \; 
           \Sigma:(3,0,1); \; \Delta:(3,0,1); \;
           H:(2,1,0);\;\varphi_1:(1,0,1);\; \varphi_2:(1,0,2)$   
           \vspace{1mm} \\
           \underline{Yukawa Lagrangian}:  $-\mathscr{L}_Y =-\mathscr{L}_{Y_1} + g_{\Delta} \Tr{[\overline{\Sigma} \Delta
           \nu_R]} + g_\Sigma \Tr{[\varphi_2 \overline{\Sigma^ c}\Sigma]} + h.c.$
           \vspace{1mm} \\
           \underline{Potential}: $V \supset \lambda \varphi_1 H^Ti\sigma_2\boldsymbol{\Delta}^\dagger
           \tilde{H}+ h.c.$
           \vspace{1mm} \\
           Note that we only displayed terms in the Yukawa Lagrangian and the potential that are 
           relevant for the lowest order diagram of right-handed neutrino mass generation. 
           The diagram additional to \textbf{1C} is given by
           \vspace{7mm} \\
           \begin{fmffile}{right_hidden}
             \begin{eqnarray*}
               \parbox{65mm}{
               \begin{fmfgraph*}(65,45)
                 \fmfstraight
                 \fmfleft{L} \fmflabel{$\nu_R$}{L}
                 \fmfright{R} \fmflabel{$\nu_R$}{R}
                 \fmftop{A,T1,T2,T3,T4,T5,T6,T7,B} \fmflabel{$\langle H \rangle$}{T1}
                                                       \fmflabel{$\langle \varphi_1 \rangle$}{T2}
                                                       \fmflabel{$\langle H \rangle$}{T3}
                                                       \fmflabel{$\langle H \rangle$}{T5}
                                                       \fmflabel{$\langle \varphi_1 \rangle$}{T6}
                                                       \fmflabel{$\langle H \rangle$}{T7}
                 \fmfbottom{xA,xT1,xT2,xT3,xT4,xT5,xT6,xT7,xB}
                 \fmflabel{$\langle \varphi_2 \rangle$}{xT4}
                                            
                 \fmf{fermion}{L,V1}
                 \fmf{fermion,label=$\Sigma$}{V1,V2}
                 \fmf{fermion,label=$\Sigma$}{V3,V2} 
                 \fmf{fermion}{R,V3}
              
                 \fmffreeze
              
                 \fmf{scalar,label=$\Delta$,tension=2.5}{TV1,V1}
                 \fmf{scalar,tension=2.5}{xT4,V2}
                 \fmf{scalar,label=$\Delta$,tension=2.5}{TV3,V3}
              
                 \fmf{scalar}{T1,TV1}
                 \fmf{scalar}{T2,TV1}
                 \fmf{scalar}{TV1,T3}
              
                 \fmf{scalar}{T5,TV3}
                 \fmf{scalar}{T6,TV3}
                 \fmf{scalar}{TV3,T7}
               \end{fmfgraph*}            
               }
             \end{eqnarray*}
           \end{fmffile}
           \vspace{1cm} \\
           If $\Delta$ does not get a VEV at tree level this is the leading tree-level diagram in the $3 \times 3$ space as the term
           $\varphi \overline{\nu_R} \nu_R^c$ is forbidden. The right-handed mass $M_R$ can
           therefore be estimated using the same argument as in \textbf{1B}
           \begin{equation}
             M_R = \frac{\lambda^2 g_{\Delta}^2}{g_\Sigma}
             \left( \frac{\langle H \rangle}{M_\Delta} \right)^4 
             \frac{\langle \varphi_1 \rangle^2}{\langle \varphi_2 \rangle} \lesssim \frac{ g_{\Delta}^2}{{g_\Sigma}} \frac{\text{GeV}^2}{\ev{\varphi_2}}.
           \end{equation}

           This means that the mass matrix is given by
           \begin{equation}
             \mathcal{M} = 
             \begin{pmatrix}
               0 & m_D & 0 \\
               m_D & M_R & M_{Rx} \\
               0 & M_{Rx} & M_x
             \end{pmatrix} \, .
           \end{equation}
      Which is similar to \textbf{1C} but with a non vanishing $M_R$ at tree level.

       Now we turn to a theory with different phenomenology by forbidding the VEV of $\varphi_1$.

         \item
           \underline{\textbf{5C: SM  $ \times \,\boldsymbol{U_H(1)} $}} \\
           \underline{Particle content}: $L:(2,-1,0) ;\; \nu_R:(1,0,0);\; \Sigma:(3,0,1);\; \Delta:(3,0,1);\;
           H:(2,1,0);\; \varphi_1:(1,0,1);\; \varphi_2:(1,0,2);\; \varphi_3:(1,0,-4)$
           \vspace{1mm} \\
           The Yukawa Lagrangian is the same as in the previous theory, while we get an additional
           potential term.
           \vspace{1mm} \\
           \underline{Potential}: $V \supset \lambda \varphi_1 H^Ti\sigma_2 
           \boldsymbol{\Delta}^\dagger \tilde{H}
           + \lambda' \varphi_1^2 \varphi_2 \varphi_3 + h.c. + ...$
           \vspace{1mm} \\
          With $\langle \varphi_1 \rangle=0$ the lowest order diagram 
           contributing to the right-handed neutrino mass is given by
           \vspace{7mm} \\
           \begin{fmffile}{right_hidden_radiative}
             \begin{eqnarray*}
               \parbox{65mm}{
               \begin{fmfgraph*}(65,55)
                 \fmfleft{L} \fmflabel{$\nu_R$}{L}
                 \fmfright{R} \fmflabel{$\nu_R$}{R}
                 \fmftop{A,T1,T2,B,T3,T4,C,T5,T6,D} \fmflabel{$\langle H \rangle$}{T1}
                                                    \fmflabel{$\langle H \rangle$}{T2}
                                                    \fmflabel{$\langle \varphi_2 \rangle$}{T3}
                                                    \fmflabel{$\langle \varphi_3 \rangle$}{T4}
                                                    \fmflabel{$\langle H \rangle$}{T5}
                                                    \fmflabel{$\langle H \rangle$}{T6}
                 \fmfbottom{B}  \fmflabel{$\langle \varphi_2 \rangle$}{B}   
                                                 
                 \fmf{fermion}{L,V1}
                 \fmf{fermion,lab=$\Sigma$}{V1,V2}
                 \fmf{fermion,lab=$\Sigma$}{V3,V2}
                 \fmf{fermion}{R,V3}
                 
                 \fmffreeze
                 
                 \fmf{scalar,right=1/4,tension=1.5,lab=$\Delta$}{TV1,V1}
                 \fmf{scalar,right=1/4,lab=$\varphi_1$}{TV2,TV1}
                 \fmf{scalar,left=1/4,lab=$\varphi_1$}{TV2,TV3}
                 \fmf{scalar,left=1/4,tension=1.5,lab=$\Delta$}{TV3,V3}
                 
                 \fmf{scalar,tension=.3}{T1,TV1}
                 \fmf{scalar,tension=.3}{TV1,T2}
                 \fmf{scalar,tension=.3}{TV2,T3}
                 \fmf{scalar,tension=.3}{TV2,T4}
                 \fmf{scalar,tension=.3}{T5,TV3}
                 \fmf{scalar,tension=.3}{TV3,T6}
                 
                 \fmf{scalar}{V2,B}
               \end{fmfgraph*}
               }
             \end{eqnarray*}
           \end{fmffile}
           \vspace{7mm} \\
          Using the fact that the loop has the same topology as in \textbf{3C} and just the external VEVs are different we get 
           
           \begin{align}
             M_R \approx  \frac{\lambda^2 \lambda' }{16\pi^2} \left(\frac{\ev{H}}{M_\Delta}\right)^4 \frac{ g_\Delta^2}{g_\Sigma} \ev{\varphi_3} I\left(\frac{\ev{\varphi_2}^2}{M_{\varphi_1}^2} \right)  \, .           
           \end{align}

          This loop suppression combined with a mass suppression to the fourth power with the Triplet mass can generate the Pseudo-Dirac scenario for active neutrinos without large fine tuning in the Majorana mass sector.

          \subsubsection*{Modifying the $\nu_x$ Majorana Mass}
      The general mass matrix structure for the following models will be of the form
      \begin{equation}
        \label{eq:inverse}
        \mathcal{M} = 
        \begin{pmatrix}
          0 & m_D & 0 \\
          m_D & 0 & M_{Rx} \\
          0 & M_{Rx} & M_x
        \end{pmatrix} \, .
      \end{equation}

        \item
          \underline{\textbf{1D: SM  $ \times \,\boldsymbol{U_H(1)} $}} \\
          \underline{Particle content}: $ L:(2,-1, 0) ;\; \nu_R:(1,0,0) ;\; \nu_x:(1,0,1);\; \Sigma:(3,0,-2);\; H:(2,1,0);\; 
          \varphi_1:(1,0,-3);\;\varphi_2:(1,0,-4);\; \Delta:(3,0,-3); \; \varphi_4:(1,0,1)$
          \vspace{1mm} \\
          \underline{Yukawa Lagrangian}: $-\mathscr{L}_Y \supset g_H \overline{L}\tilde{H}\nu_R +  g_{Rx} \varphi_4 \overline{\nu_R}
          \nu_x^c + g_{\Delta} \Tr{[\overline{\Sigma} \Delta
          \nu_x]} + g_\Sigma \Tr{[\varphi_2 \overline{\Sigma^ c}\Sigma]} + h.c.$
          \vspace{1mm} \\
          \underline{Potential}: $V \supset \lambda \varphi_1 H^Ti\sigma_2\boldsymbol{\Delta}^\dagger
          \tilde{H}+ h.c. + ...$
          \vspace{1mm} \\
         The leading diagram is
          \vspace{7mm} \\
          \begin{fmffile}{x_hidden}
            \begin{eqnarray*}
              \parbox{65mm}{
              \begin{fmfgraph*}(65,45)
                \fmfstraight
                \fmfleft{L} \fmflabel{$\nu_x$}{L}
                \fmfright{R} \fmflabel{$\nu_x$}{R}
                \fmftop{A,T1,T2,T3,T4,T5,T6,T7,B} \fmflabel{$\langle H \rangle$}{T1}
                                                      \fmflabel{$\langle \varphi_1 \rangle$}{T2}
                                                      \fmflabel{$\langle H \rangle$}{T3}
                                                      \fmflabel{$\langle H \rangle$}{T5}
                                                      \fmflabel{$\langle \varphi_1 \rangle$}{T6}
                                                      \fmflabel{$\langle H \rangle$}{T7}
                \fmfbottom{xA,xT1,xT2,xT3,xT4,xT5,xT6,xT7,xB}
                \fmflabel{$\langle \varphi_2 \rangle$}{xT4}
                                            
                \fmf{fermion}{L,V1}
                \fmf{fermion,label=$\Sigma$}{V1,V2}
                \fmf{fermion,label=$\Sigma$}{V3,V2} 
                \fmf{fermion}{R,V3}
              
                \fmffreeze
              
                \fmf{scalar,label=$\Delta$,tension=2.5}{TV1,V1}
                \fmf{scalar,tension=2.5}{xT4,V2}
                \fmf{scalar,label=$\Delta$,tension=2.5}{TV3,V3}
              
                \fmf{scalar}{T1,TV1}
                \fmf{scalar}{T2,TV1}
                \fmf{scalar}{TV1,T3}
              
                \fmf{scalar}{T5,TV3}
                \fmf{scalar}{T6,TV3}
                \fmf{scalar}{TV3,T7}
              \end{fmfgraph*}            
              }
            \end{eqnarray*}
          \end{fmffile}
          \vspace{0.5cm} \\

      The mass matrix is given by eq. (\ref{eq:inverse}) and the Majorana mass of $\nu_x$ is 
          \[
            M_x = \frac{\lambda^2 g_{\Delta}^2}{g_\Sigma}
            \left( \frac{\langle H \rangle}{M_\Delta} \right)^4 
            \frac{\langle \varphi_1 \rangle^2}{\langle \varphi_2 \rangle} \lesssim \frac{ g_{\Delta}^2}{{g_\Sigma}} \frac{\text{GeV}^2}{\ev{\varphi_2}} ,     
          \]
          where the suppression of the small lepton number violating contribution by the heavy scalar VEV makes it an inverse see-saw scenario. Implying sterile neutrinos with at the TeV scale and slightly above. Those form pseudo Dirac pairs and can have seizable mixing with the active neutrinos. 
With the mass scale of $\ev{\varphi_2}$ around a few TeV and the yuakawa couplings of $g_\Delta \approx 10^{-1}$ and $ g_\Sigma \approx 1$, the scale $M_x$ is naturally at the keV scale, which is required phenomenologically to have sub eV active neutrino masses. The active-sterile mixing is approximately given by $(m_D/M_{Rx})^2$ and can in principle range from 1$\%$ to undetectable values below $10^{-10}$.  The interesting observation is that small active-sterile mixing requires unnaturaly small Dirac Yukawa couplings in this model.

      It is possible that the $\nu_x$ Majorana masses are generated radiatively.  
        \item
          \underline{\textbf{2D: SM  $ \times \,\boldsymbol{U_H(1)} $}} \\
          \underline{Particle content}: $ L:(2,-1, 0) ;\; \nu_R:(1,0,0) ;\; \nu_x:(1,0,1);\; \Sigma:(3,0,-2);\; H:(2,1,0);\; 
          \varphi_1:(1,0,-3);\; \varphi_2:(1,0,-4);\; \varphi_3:(1,0,10)\Delta:(3,0,-3);\; \varphi_4:(1,0,1)$
          \vspace{1mm} \\
          Here $\ev{\varphi_1}=0$ and the Yukawa Lagrangian
          is the same as in the theory before. The potential, however, is extended.
          \vspace{1mm} \\
          \underline{Potential}: $V \supset \lambda \varphi_1 H^Ti\sigma_2\boldsymbol{\Delta}^\dagger
          \tilde{H} + \lambda' \varphi_1^2 \varphi_2 \varphi_3 + h.c.$
          \vspace{1mm} \\
          We obtain the following diagram
          \vspace{7mm} \\
          \begin{fmffile}{x_hidden_radiative}
            \begin{eqnarray*}
              \parbox{65mm}{
              \begin{fmfgraph*}(65,55)
                \fmfleft{L} \fmflabel{$\nu_x$}{L}
                \fmfright{R} \fmflabel{$\nu_x$}{R}
                \fmftop{A,T1,T2,B,T3,T4,C,T5,T6,D} \fmflabel{$\langle H \rangle$}{T1}
                                                   \fmflabel{$\langle H \rangle$}{T2}
                                                   \fmflabel{$\langle \varphi_2 \rangle$}{T3}
                                                   \fmflabel{$\langle \varphi_3 \rangle$}{T4}
                                                   \fmflabel{$\langle H \rangle$}{T5}
                                                   \fmflabel{$\langle H \rangle$}{T6}
                \fmfbottom{B} \fmflabel{$\langle \varphi_2 \rangle$}{B}
                                                    
                \fmf{fermion}{L,V1}
                \fmf{fermion,lab=$\Sigma$}{V1,V2}
                \fmf{fermion,lab=$\Sigma$}{V3,V2}
                \fmf{fermion}{R,V3}
                
                \fmffreeze
                
                \fmf{scalar,right=1/4,tension=1.5,lab=$\Delta$}{TV1,V1}
                \fmf{scalar,right=1/4,lab=$\varphi_1$}{TV2,TV1}
                \fmf{scalar,left=1/4,lab=$\varphi_1$}{TV2,TV3}
                \fmf{scalar,left=1/4,tension=1.5,lab=$\Delta$}{TV3,V3}
                
                \fmf{scalar,tension=.3}{T1,TV1}
                \fmf{scalar,tension=.3}{TV1,T2}
                \fmf{scalar,tension=.3}{TV2,T3}
                \fmf{scalar,tension=.3}{TV2,T4}
                \fmf{scalar,tension=.3}{T5,TV3}
                \fmf{scalar,tension=.3}{TV3,T6}
                
                \fmf{scalar}{V2,B}
              \end{fmfgraph*}
              }
            \end{eqnarray*}
          \end{fmffile}
          \vspace{8mm} \\
          As before the Majorana mass of $\nu_x$ can be approximated by
          \begin{equation}
            M_R \sim 10^{-2} \cdot \frac{g_\Delta^2 \lambda^2 \lambda'}{g_\Sigma}
            \left( \frac{\ev{H}}{M_\Delta} \right)^4\cdot EWS \, .
          \end{equation}
          We see that in this setup the $\nu_x$ mass is at the keV scale when the Yukawa couplings are of order one, the potential terms between 0.1 and one and the Triplet around the TeV scale. This is the right scale for the inverse see-saw scenario. Note that as before we need another scalar $\varphi_4$ for 
          the connection between SM sector and Hidden Sector.

\section{\label{app:Exceptions} Fully Radiative Generated Left-Handed Masses}
 		
			As was shown by \ref{sec:Radiative} there is no way of generating left-handed neutrino masses radiatively by pairwise coupling scalars in the potential.
			We go through the five possibilities for non pairwise coupling of scalars and study whether radiative mass generation is possible. Furthermore, we present possibilities to circumvent \ref{sec:Radiative}.
		
		 \item 
            \textbf{Possibility 1}: We can introduce a potential coupling of four different $SU(2)$
            singlet scalars such that their hypercharges add up to zero. In this case one $SU(2)$
            singlet with vanishing hypercharge has to be included as we need an electrically
            neutral scalar to gain a VEV. \\
            With this kind of coupling it is indeed possible to construct a theory that generates
            neutrino masses fully radiatively. Consider as an example the  theory \textbf{11A}. \\
            \underline{Particle content}: $L:(2,-1);\; \ell_R:(1,-2);\; H:(2,1);\;\delta_-:(1,-2);\;
            \epsilon_{++}:(1,4)$ 
            \vspace{1mm} \\
            \underline{Yukawa Lagrangian}:
            $-\mathscr{L}_Y \supset g_\delta\bar{L}L^c\delta_- + g_\epsilon \overline{l_R^c} \ell^{}_R
            \epsilon_{++} + \overline{L} H \ell_R + h.c.$
            \vspace{1mm} \\
            \underline{Potential}: $V \supset \lambda\varphi \delta_- \delta_- \epsilon_{++} + h.c. + ... $
            \vspace{1mm} \\
            For this theory we find the radiative generation of neutrino masses represented by the
            diagram:
             \\
            \begin{fmffile}{Zee_conf}
              \begin{eqnarray*}
                \parbox{75mm}{
                \begin{fmfgraph*}(75,55)
                  \fmfstraight
             
                  \fmfleft{L} \fmflabel{L}{L}
                  \fmfright{R} \fmflabel{L}{R}
                  \fmftop{P1,P2,P3} \fmflabel{$\langle \varphi \rangle$}{P2}

                  \fmf{fermion,tension=1}{L,V1}
                  \fmf{fermion,label=L}{V2,V1}
                  \fmf{fermion,label=$l_R$}{V3,V2}
                  \fmf{fermion,label=$l_R$}{V3,V4}
                  \fmf{fermion,label=L}{V4,V5}
                  \fmf{fermion,tension=1}{R,V5}
               
                  \fmffreeze
               
                  \fmf{scalar,left=.3,label=$\delta_-$}{V1,VT}
                  \fmf{scalar,right=.3,label=$\delta_-$}{V5,VT}
                  \fmf{phantom,tension=2.5}{P1,VT}
                  \fmf{scalar,tension=.5}{P2,VT}
                  \fmf{phantom,tension=2.5}{P3,VT}   
               
                  \fmf{scalar,tension=3,label=$\epsilon_{++}$}{V3,VT}
               
                  \fmfv{decor.shape=cross,decor.size=8}{V2}
                  \fmfv{decor.shape=cross,decor.size=8}{V4}
                \end{fmfgraph*}             
                }
              \end{eqnarray*}
            \end{fmffile}
            \vspace{-2cm} \\
            The crosses denote the insertion of a Higgs VEV, i.e. they represent the mass of the
            charged lepton. This theory is the conformally invariant analogue to the Zee-Babu model.
            The corresponding left-handed neutrino mass is given by
            \begin{equation}
              M_L = 8 \lambda \langle \varphi \rangle m_l^2 g_\delta^2 g_\epsilon I \, ,
            \end{equation}
            where $I$ is given by eq. (\ref{eq:Integral}).
            \begin{equation}
         \label{eq:Integral}
         \begin{split}
           I = \int \! \frac{\mathrm{d}^4p}{(2\pi)^4} \; \int \! \frac{\mathrm{d}^4q}{(2\pi)^4} \;
           \frac{1}{p^2-m_l^2} \frac{1}{q^2-m_l^2} \\
           \frac{1}{p^2-m_\delta^2} \frac{1}{q^2-m_\delta^2} \frac{1}{(p-q)^2-m_\epsilon^2} \, .
         \end{split}
         \end{equation}
            
		  \item
		   \textbf{Possibility 2}: We can introduce an additional $SU(2)$ doublet $H_2$, an additional charged scalar singlet $\eta_+$ and a total singlet $\varphi: (1,0)$.  As stated by before there has to be a term in the potential with non pairwise coupled scalars. This $\lambda_L$ term violates lepton number and its size controls the neutrino masses.
		   
		   \underline{Particle content}: $L:(2,-1);\; \ell_R:(1,-2);\; H_1:(2,1);\ H_2:(2,1);\;\eta_+:(1,+2);\;
            \varphi:(1,0)$ 
            \vspace{1mm} \\
            With additional terms in the Yukawa Lagrangian and potential.
            
            \underline{Yukawa Lagrangian}:
            $-\mathscr{L}_Y \supset  g_1\, \eta_+ \bar{L}\,i \sigma_2 L^c + g_2\,\bar{L} H_2 \ell_R + h.c.$
            \vspace{1mm} \\
            \underline{Potential}: $V \supset \lambda_L \eta \tilde{H}_1^\dagger H_2 \, \varphi  + h.c. + ... $
            \vspace{1mm} \\
            The loop diagram gives neutrino masses
		   \vspace{7mm} \\
           \begin{fmffile}{Zee_radiative}
             \begin{eqnarray*}
               \parbox{55mm}{
               \begin{fmfgraph*}(55,45)
                 \fmfleft{L} \fmflabel{$L$}{L}
                 \fmfright{R} \fmflabel{$L$}{R}
                 \fmftop{TA,T1,T2,TB} \fmflabel{$\langle H_{2/1} \rangle$}{T1}
                                    \fmflabel{$\langle \varphi \rangle$}{T2}
                 \fmfbottom{BA,BB,B,BC,BD} 
                 
                 \fmf{fermion}{L,V1}
                 \fmf{fermion,label=$\ell_R$}{V2,V1}
                 \fmf{fermion,label=$L$}{V2,V3}
                 \fmf{fermion}{R,V3}
                 
                 \fmffreeze
                 
                 \fmf{scalar,left=.5,label=$H_{1/2}$}{V1,TV}
                 \fmf{scalar,right=.5,label=$\eta_+$}{V3,TV}
                 \fmf{scalar,tension=1.3}{T1,TV}
                 \fmf{scalar,tension=1.3}{T2,TV}
                \fmfv{decor.shape=cross,decor.size=8}{V2}
               \end{fmfgraph*}
               }
             \end{eqnarray*}
           \end{fmffile}
           \vspace{-8mm} \\
           which have the mass pattern as the non conformal Zee model \cite{Zee:1985rj}, with the difference that the dimensionful parameter controlling the neutrino masses is replaced by the product of the coupling with the scalar VEV $\lambda_L \cdot \ev{\varphi}$, see \textbf{12A}.
            
          \item 
            \textbf{Possibility 3}: We can introduce a potential coupling of 3 different $SU(2)$
            doublets such that their hypercharges add up to zero in the following structure
            \begin{equation}
              \left( \phi_1^\dagger \vec{\sigma} H_i \right) \left( \tilde{H}_j^\dagger \vec{\sigma} H_j \right) \, .
            \end{equation}               
           	As proposed in \cite{Law:2013dya} an additional doubly charged singlet scalar can be used to gain neutrino masses at two loop level. In a conformal 			model, however, an additional scalar is required to have a lepton number violating term in the Lagrangian without an explicit mass scale.   
            
            \underline{Particle content}: $ L:(2,-1);\; \ell_R (1 ,-2) ;\; \phi_1:(2,3);\; H_1:(2,1);\; H_2:(2,1) ;\;  \eta:(1,-4) ;\; \phi_2:(1,0)$
           \vspace{+0.5mm} \\ 
  			\underline{Yukawa Lagrangian}:
            $-\mathscr{L}_Y \supset  g\, \eta \, \bar{\ell}_R \ell_R^c + h.c. $
            \vspace{+0.5mm} \\            
            \underline{Potential}: 
            
            $ V \supset \lambda_i \,\phi_2 \eta \phi_1^\dagger \tilde{H}_i  + \lambda_{ij}  \left( \phi_1^\dagger \vec{\sigma} H_i \right) \left( 				  \tilde{H}_j^\dagger \vec{\sigma} H_j \right) $
            
           Here, both doublets $H_1$ and $H_2$ as well as the singlet scalar get a VEV and generate neutrino masses at two loop level.
              
          \item
            \textbf{Possibility 4}: A potential term coupling 4 different $SU(2)$ triplets such that
            their hypercharges add up to zero in the following way
            \begin{equation}
              \left( \Delta_1^\dagger \Delta_2 \right) \left( \Delta_3^\dagger \Delta_4 \right) \, .
            \end{equation}
           This term generates neutrino masses at the two loop level with the same topology as in the conformal Zee-Babu model in example 1.

          \item
            \textbf{Possibility 5}: A further term that can be introduced is given by the coupling
            \begin{equation}
              \varphi H_1^Ti\sigma_2\boldsymbol{\Delta}^\dagger H_2 \, ,
            \end{equation}             
            where $\varphi$ is a $SU(2)$ singlet, $H_1$ and $H_2$ are doublets and $\Delta$ is
            a $SU(2)$ triplet with hypercharges such that they add up to zero in this term.
            That with the help of such a coupling the fully radiative generation of neutrino masses
            is possible can be seen in the following theory: 
            \vspace{1mm}\\
            \underline{Particle content}: $L_1:(2,-1);\; L_2:(2,-3);\; L_3:(2,0)$
            \vspace{1mm} \\ 
            \phantom{\underline{Particle content}:} $\Delta_1:(3,-4);\;
            \Delta_2:(3,-3);\; \Delta_3:(3,-1)$\\ 
            \phantom{\underline{Particle content}:} $H_1:(2,1);\; H_2:(2,-1);\; H_3:(2,-3);\; 
            H_4:(2,0)$\\
            \phantom{\underline{Particle content in loop}:} $\varphi:(1,0)$ \\
            \underline{Yukawa Lagrangian}: $-\mathcal{L}_Y \supset g_a\bar{L}_1\vec{\sigma}\Delta_1 L^c_2
            + g_b\bar{L}_2\vec{\sigma}\Delta_2 L^c_3 + g_c\bar{L}_3\vec{\sigma}\Delta_3 L^c_1 
            + h.c.$
            \vspace{.01mm} \\
            \underline{Potential}: 
            
              \begin{align}
              &\nonumber V = \lambda_a\varphi H_2^Ti\sigma_2 
              \boldsymbol{\Delta}_1^\dagger H_3\\ \nonumber
              &+ \lambda_b\varphi H_3^Ti\sigma_2\boldsymbol{\Delta}_2^\dagger H_4
              + \lambda_c\varphi H_4^Ti\sigma_2\boldsymbol{\Delta}_3^\dagger H_2 + h.c. \\ \nonumber
              &+ \lambda_{13}(H_3^\dagger H_3)(H_1^\dagger H_1) + \lambda_{14}(H_4^\dagger H_4)
              (H_1^\dagger H_1) \\ \nonumber
              &+ \text{pairwise couplings}
              \end{align}
            
            If we forbid the VEVs $\langle \Delta_1 \rangle$, $\langle \Delta_2 \rangle$ and
            $\langle \Delta_3 \rangle$, then
            the following diagram describes the radiative generation of neutrino masses:
            \vspace{3mm} \\
            \begin{fmffile}{scalar}
              \begin{eqnarray*}
                \parbox{65mm}
                {
                \begin{fmfgraph*}(65,55)
                  \fmfleft{L1} \fmflabel{$L_1$}{L1}
                  \fmfright{L2} \fmflabel{$L_1$}{L2}
                  \fmftop{A,S1,G,S2,C,E,S3,D,F,S4,H,S5,B} 
                  \fmflabel{$\langle\varphi\rangle$}{S1}
                  \fmflabel{$\langle H_2 \rangle$}{S2}
                  \fmflabel{$\langle\varphi\rangle$}{S3}
                  \fmflabel{$\langle H_2 \rangle$}{S4}
                  \fmflabel{$\langle\varphi\rangle$}{S5}
       
                  \fmf{fermion,tension=2}{L1,I1}
                  \fmf{fermion,label=$L_2$}{I2,I1}
                  \fmf{fermion,label=$L_3$}{I2,I3}
                  \fmf{fermion,tension=2}{L2,I3}
     
                  \fmffreeze
     
                  \fmf{scalar,left=1/4,label=$\Delta_1$,tension=4.3}{I1,V1}
                  \fmf{scalar,left=1/4,label=$H_3$,tension=4.3}{V1,V2}
                  \fmf{scalar,right=1/4,label=$H_4$,tension=4.3}{V3,V2}
                  \fmf{scalar,right=1/4,label=$\Delta_3$,tension=4.3}{I3,V3}
                  \fmf{scalar,label=$\Delta_2$,tension=.1}{V2,I2}
     
                  \fmf{scalar,tension=1.7}{V1,S1}
                  \fmf{scalar,tension=1.7}{V1,S2}
                  \fmf{scalar,tension=5}{S3,V2}
                  \fmf{scalar,tension=1.7}{V3,S4}
                  \fmf{scalar,tension=1.7}{S5,V3}
                \end{fmfgraph*}
                }
              \end{eqnarray*}
            \end{fmffile}
            \vspace{-2cm} \\
            Admittedly this theory is very baroque and can be phenomenologically problematic. Especially to ensure anomaly cancellation the new fermions have to be vector like. The particle content in the loop is intended to show that it is possible to generate neutrino masses fully radiatively from a topological point of view.

          \item
            \textbf{Alternative 1}: So far in the radiative models no additional symmetries were considered. However, the argument of \ref{sec:Radiative} can be avoided if a new symmetry is present, which forbids tree level couplings for fermion singlets in the SM Dirac term. If there is a discrete symmetry, for example $Z_2$ under  which all SM particles are even and the spectrum given by
            
			\underline{Particle content}: $ L(2,1, (+)) ;\; H_1:(2,1, (+));\; H_2:(2,1, (-));\; \nu_x:(1,0,(-));\;\varphi:(1,0,(+))$\\
          \vspace{-3mm} \\ 
            Additional in the Yuakawa Lagrangian there are the following terms.
            
            \underline{Yukawa Lagrangian}: $-\mathscr{L}_Y \supset g_{1}\bar{L} H_2 \nu_x + g_2 \varphi \bar{\nu_x} \nu_x^c  + h.c.$ 
             
             The relevant coupling in the potential is:      
             
             \underline{Potential}:  $V \supset \lambda\, (H_2^\dagger H_1)^2 + \text{pairwise couplings}$
             
            with $H_1$ being the SM Higgs. This would be the conformal analogue of the Ma model and generates neutrino masses at one loop level. Note, however, that in general discrete symmetries are not so restrictive. Therefore, in our models continuous symmetries are used. For example a hidden sector $U(1)$ would have the same effect on the Yukawa Lagrangian, but the potential term would be forbidden. Thus a model of this type can only generate neutrino masses in the hidden sector, as shown in the model \textbf{3C}.

          \item
            \textbf{Alternative 2}: To circumvent the argument of \ref{sec:Radiative} we can allow fermion loops. The following theory 
            shows that it is possible to construct left-handed Majorana masses such that the 
            lowest order diagram has to be a full loop diagram.
            \vspace{1mm} \\
            \underline{Particle content}: $L_1:(2,-1);\; L_2:(2,0);\; L_3:(2,2);\; L_4:(2,-2)$\\
            \phantom{\underline{Particle content}:} $\Delta_1:(3,-2);\; \Delta_2:(3,0);\; 
            	\Delta_3:(3,2)$
            \vspace{1mm} \\
            \underline{Yukawa Lagrangian}: $-\mathscr{L}_Y \supset g_{11}\bar{L}_1\vec{\sigma}\Delta_1 L_1^c
            + g_{24}\bar{L}_2\vec{\sigma}\Delta_1 L_4^c + g_{22}\bar{L}_2\vec{\sigma}\Delta_2 L_2^c
            \\ \phantom{\underline{Yukawa Lagrangian}: -\mathcal{L}_Y=}
            + g_{23}\bar{L}_2\vec{\sigma}\Delta_3 L_3^c + g_{34}\bar{L}_3\vec{\sigma}\Delta_2 L_4^c 
            + h.c.$ 
            \vspace{1mm} \\
            Furthermore we require the VEV of the neutral component of $\Delta_1$ to vanish.
            The following diagram is then the lowest order contribution to the left-handed neutrino 
            masses:
            \vspace{1cm} \\
            \begin{fmffile}{fermion}
              \begin{eqnarray*}
                \parbox{35mm}
                {
                \begin{fmfgraph*}(35,85)
                  \fmfleft{L1} \fmflabel{$L_1$}{L1}
                  \fmfright{L2} \fmflabel{$L_1$}{L2}
                  \fmftop{S1,S2,S3} 
                    \fmflabel{$\langle\Delta_2\rangle$}{S1}
                    \fmflabel{$\langle\Delta_3\rangle$}{S2}
                    \fmflabel{$\langle\Delta_2\rangle$}{S3}
                  \fmfbottom{P1,P2,P3}
       
                  \fmf{fermion,tension=3}{L1,I1}
                  \fmf{fermion,tension=3}{L2,I1}
                  
                  \fmffreeze
       
                  \fmf{scalar,tension=3.5,label=$\Delta_1$}{I1,V1}  
       
                  \fmf{fermion,left=1/3,tension=3,label=$L_2$}{V1,V2}    
                  \fmf{fermion,right=1/6,label=$L_2$}{V3,V2}
                  \fmf{fermion,left=1/6,label=$L_3$}{V3,V4}
                  \fmf{fermion,right=1/3,tension=3,label=$L_4$}{V1,V4}
       
                  \fmf{scalar,tension=3}{V2,S1}
                  \fmf{scalar}{S2,V3}
                  \fmf{scalar,tension=3}{V4,S3}
       
                \end{fmfgraph*}
                }
              \end{eqnarray*}
            \end{fmffile}
             \\
            Like before this theory can be phenomenologically problematic. And again it is only intended to show 
            that the topological possibility of fully radiative mass generation in
            conformally invariant theories with pairwise scalar coupling exists when introducing fermion loops.

        \end{itemize}

\bibliographystyle{apsrev}
\bibliography{references}

\end{document}